\documentclass[12pt]{article}
\usepackage{amssymb,amsmath,epsfig}
\allowdisplaybreaks

\begin{document}

\title{\bf Isotropization and Complexity Analysis of Decoupled Solutions in $f(\mathbb{R},\mathbb{T})$ Theory}
\author{M. Sharif$^1$ \thanks{msharif.math@pu.edu.pk} and Tayyab Naseer$^{1,2}$ \thanks{tayyabnaseer48@yahoo.com}\\
$^1$ Department of Mathematics and Statistics, The University of Lahore,\\
1-KM Defence Road Lahore, Pakistan.\\
$^2$ Department of Mathematics, University of the Punjab,\\
Quaid-i-Azam Campus, Lahore-54590, Pakistan.}

\date{}
\maketitle

\begin{abstract}
This paper formulates some new exact solutions to the field
equations by means of minimal gravitational decoupling in the
context of $f(\mathbb{R},\mathbb{T})$ gravity. For this purpose, we
consider anisotropic spherical matter distribution and add an extra
source to extend the existing solutions. We apply the transformation
only on the radial metric potential that results in two different
sets of the modified field equations, each of them corresponding to
their parent source. The initial anisotropic source is represented
by the first set, and we consider two different well-behaved
solutions to close that system. On the other hand, we impose
constraints on the additional source to make the second set
solvable. We, firstly, employ the isotropization condition which
leads to an isotropic system for a particular value of the
decoupling parameter. We then use the condition of zero complexity
of the total configuration to obtain the other solution. The
unknowns are determined by smoothly matching the interior and
exterior spacetimes at the hypersurface. The physical viability and
stability of the obtained solutions is analyzed by using the mass
and radius of a compact star $4U 1820-30$. It is concluded that both
of our extended solutions meet all the physical requirements for
considered values of the coupling/decoupling parameters.
\end{abstract}
{\bf Keywords:} $f(\mathbb{R},\mathbb{T})$ gravity;
Anisotropy; Gravitational decoupling; Self-gravitating systems. \\
{\bf PACS:} 04.50.Kd ; 04.40.Dg;  04.40.-b.

\section{Introduction}

Cosmological discoveries show that the astronomical structures are
not distributed randomly in the universe but are arranged in a
systematic way. The investigation of such an organized pattern and
physical features of interstellar objects enable us to uncover the
cosmic accelerated expansion. In order to explain this expansion,
several modifications in general relativity ($\mathbb{GR}$) were
suggested. The first extension is $f(\mathbb{R})$ theory obtained by
replacing the Ricci scalar $\mathbb{R}$ with its generic function in
an Einstein-Hilbert action. A large body of literature exists to
discuss this theory as the first attempt to explain inflationary as
well as present (accelerated expansion) epochs \cite{2,2a}. Multiple
techniques have been employed to analyze the stability of this
extended theory \cite{9,9g}.

The effect of coupling between matter distribution and geometry in
$f(\mathbb{R})$ theory was initially studied by Bertolami et al.
\cite{10} by taking the Lagrangian as a function of $\mathbb{R}$ and
$\mathbb{L}_{m}$. Such couplings encouraged astronomers to put their
attention in discussing accelerated expansion of the cosmos. Harko
et al. \cite{20} introduced this interaction on the action level by
proposing $f(\mathbb{R},\mathbb{T})$ theory, in which $\mathbb{T}$
refers to trace of the energy-momentum tensor ($\mathbb{EMT}$). The
non-conservation of the $\mathbb{EMT}$ has been observed in this
theory and an extra force is always present (depending on density
and pressure \cite{22}) that helps the test particles to move in
non-geodesic path. This theory comprises new gravitational aspects
due to the inclusion of $\mathbb{T}$ and also successfully meets
weak-field solar system conditions. Houndjo \cite{22a} explained how
the matter-dominated era switches into late-time acceleration phase
by employing minimally coupled $f(\mathbb{R},\mathbb{T})$ model. A
particular model such as $\mathbb{R}+2\varpi\mathbb{T}$ has become
very popular among the researchers during the last few years. Das et
al. \cite{22b} used this model to discuss structure of the
three-layers gravastar, each of these sectors is expressed by its
corresponding equation of state. The interior of various stellar
systems has been discussed in this modified scenario \cite{25a,25b}.

The gravitational field equations representing stellar models
incorporate highly non-linear terms that always prompt
astrophysicists to think how to find their exact solutions. The
formulation of well-behaved solutions of such equations may prove
useful to know the nature of realistic physical systems. For this
reason, researchers put their efforts to produce viable cosmic
objects with the help of multiple methods. A recently developed
technique is the gravitational decoupling used to develop feasible
solutions analogous to the stellar bodies whose interior may be
filled with different sources (such as anisotropy, shear and
dissipation flux). This technique helps to solve the field equations
involving more than one matter source by decoupling them into
multiple sets, each of them correspond to their parent source.
Ovalle \cite{29} developed the minimal geometric deformation (MGD)
approach for the very first time which offers a class of appealing
ingredients for exact solutions of stellar objects in the braneworld
scenario. Later, Ovalle and Linares \cite{30} discussed spherical
source coupled with an isotropic configuration to formulate the
corresponding solution that was observed to be compatible with the
Tolman-IV ansatz in the braneworld. Casadio et al. \cite{31}
extended this technique in the Randall-Sundrum braneworld and
obtained the corresponding Schwarzschild geometry.

Ovalle et al. \cite{33} used the MGD approach to develop an
extension of isotropic source to the new anisotropic source and
analyzed graphical behavior of the resulting solutions. Sharif and
Sadiq \cite{34} constructed two different decoupled solutions for
charged anisotropic sphere by employing the Krori-Barua ansatz as an
isotropic solution and observed the impact of electromagnetic field
on their stability. This approach has been used in $f(\mathbb{G})$
and $f(\mathbb{R})$ gravitational theories and several anisotropic
solution were obtained \cite{35}. Gabbanelli et al. \cite{36}
extended an isotropic Duragpal-Fuloria ansatz to formulate different
physically acceptable anisotropic solutions. The extension of the
Heintzmann solution to multiple stable anisotropic decoupled
solutions has also been done \cite{36a}. The Tolman VII isotropic
ansatz has been deformed into acceptable anisotropic solutions by
Hensh and Stuchlik \cite{37a}. Sharif and Ama-Tul-Mughani \cite{37b}
found different solutions corresponding to charged string cloud as
well as uncharged axially symmetric spacetime. Several isotropic
solutions were chosen to determine anisotropic solutions by means of
minimal/extended decoupling scheme in the context of Brans-Dicke
theory \cite{37c}. We have formulated some anisotropic solutions
corresponding to different constraints in a non-minimally
matter-geometry coupled gravity \cite{37f}.

Recently, Herrera et al. \cite{37g,37h} gave the idea of complexity
in stellar systems for static as well as dynamical spherical matter
sources. They found some scalar factors from orthogonal
decomposition of the curvature tensor which inherently connect
energy density inhomogeneity, local anisotropy and pressure
components. This phenomenon has been extended for static and
non-static self-gravitating structures in modified scenario
\cite{37i,37ia}. The set of field equations can be closed by using
additional constraints such as the vanishing complexity factor or
two systems with the same complexity. The decoupling scheme along
with the above condition has widely been used to develop feasible
stellar models \cite{37j}. Casadio et al. \cite{37k} employed the
MGD technique and isotropize the anisotropic system for a particular
value of the decoupling parameter. Maurya et al. \cite{37l} explored
how the decoupling parameter affects the complexity and anisotropy
of spherical embedding class one geometry. Sharif and Majid
\cite{37m} extended this work to the Brans-Dicke theory and obtained
two stable solutions.

This paper investigates how the decoupling through MGD affects
various physical characteristics of static spherical structure in
the background of $f(\mathbb{R},\mathbb{T})$ theory. We obtain two
solutions, one is from isotropization of the anisotropic source and
other is governed by the complexity of considered setup. The paper
is organized as follows. We define some basic formulation of
modified theory and develop the field equations in the presence of
an additional source in the next section. The field equations are
then separated through MGD technique in section \textbf{3}. Two new
exact solutions are obtained in sections \textbf{4} and \textbf{5}
by employing different constraints. Section \textbf{6} discusses
physical properties of both the resulting solutions. We finally sum
up our results in section \textbf{7}.

\section{The $f(\mathbb{R},\mathbb{T})$ Theory}

The modified form of an Einstein-Hilbert action (with $\kappa=8\pi$)
in the presence of an additional field becomes \cite{20}
\begin{equation}\label{g1}
S=\int \frac{1}{16\pi}\left[f(\mathbb{R},\mathbb{T})
+\mathbb{L}_{m}+\alpha\mathbb{L}_{\mathfrak{A}}\right]\sqrt{-g}d^{4}x,
\end{equation}
where $\mathbb{L}_{m}$ and $\mathbb{L}_{\mathfrak{A}}$ are the
Lagrangian densities of fluid configuration and the additional
source which is gravitationally coupled to the matter field,
respectively. Also, $g$ describes determinant of the metric tensor
($g_{\zeta\eta}$). Taking variation of the above action with respect
to $g_{\zeta\eta}$, we have the following field equations
\begin{equation}\label{g2}
\mathbb{G}_{\zeta\eta}=8\pi \mathbb{T}_{\zeta\eta}^{(tot)},
\end{equation}
where $\mathbb{G}_{\zeta\eta}$ describes the geometric part and is
named as an Einstein tensor while the right hand side characterizes
the matter distribution as
\begin{equation}\label{g3}
\mathbb{T}_{\zeta\eta}^{(tot)}=\mathbb{T}_{\zeta\eta}^{(eff)}+\alpha
\mathfrak{A}_{\zeta\eta}=\frac{1}{f_{\mathbb{R}}}\mathbb{T}_{\zeta\eta}+\mathbb{T}_{\zeta\eta}^{(D)}+\alpha
\mathfrak{A}_{\zeta\eta}.
\end{equation}
Here, $\alpha$ is the decoupling parameter which controls the
influence of extra source ($\mathfrak{A}_{\zeta\eta}$) on
self-gravitating structure. The effective $\mathbb{EMT}$
corresponding to $f(\mathbb{R},\mathbb{T})$ gravity is represented
by $\mathbb{T}_{\zeta\eta}^{(eff)}$ and the matter source
($\mathbb{T}_{\zeta\eta}^{(D)}$) appear due to the extended gravity
has the form
\begin{eqnarray}
\nonumber \mathbb{T}_{\zeta\eta}^{(D)}&=&\frac{1}{8\pi
f_{\mathbb{R}}} \bigg[f_{\mathbb{T}}\mathbb{T}_{\zeta\eta}
+\bigg\{\frac{\mathbb{R}}{2}\bigg(\frac{f}{\mathbb{R}}-f_{\mathbb{R}}\bigg)-\mathbb{L}_{m}f_{\mathbb{T}}\bigg\}g_{\zeta\eta}\\\label{g4}
&-&(g_{\zeta\eta}\Box-\nabla_{\zeta}\nabla_{\eta})f_{\mathbb{R}}+2f_{\mathbb{T}}g^{\rho\beta}\frac{\partial^2
\mathbb{L}_{m}}{\partial g^{\zeta\eta}\partial g^{\rho\beta}}\bigg],
\end{eqnarray}
where $f_{\mathbb{R}}=\frac{\partial
f(\mathbb{R},\mathbb{T})}{\partial \mathbb{R}}$ and
$f_{\mathbb{T}}=\frac{\partial f(\mathbb{R},\mathbb{T})}{\partial
\mathbb{T}}$, $\nabla_\zeta$ and $\Box\equiv
\frac{1}{\sqrt{-g}}\partial_\zeta\big(\sqrt{-g}g^{\zeta\eta}\partial_{\eta}\big)$
are the covariant derivative and the D'Alembert operator,
respectively.

We assume the nature of seed matter source to be anisotropic whose
$\mathbb{EMT}$ can be represented as
\begin{equation}\label{g5}
\mathbb{T}_{\zeta\eta}=(\mu+P_\bot)\mathcal{K}_{\zeta}\mathcal{K}_{\eta}+P_\bot
g_{\zeta\eta}+\left(P_r-P_\bot\right)\mathcal{W}_\zeta\mathcal{W}_\eta,
\end{equation}
where $\mu,~P_r,~P_\bot,~\mathcal{K}_{\zeta}$ and
$\mathcal{W}_{\zeta}$ are termed as the energy density, radial
pressure, tangential pressure, four-velocity and the four-vector,
respectively. The trace of $f(\mathbb{R},\mathbb{T})$ field
equations can be established as
\begin{align}\nonumber
&3\nabla^{\zeta}\nabla_{\zeta}f_\mathbb{R}+\mathbb{R}f_\mathbb{R}-\mathbb{T}(f_\mathbb{T}+1)-2f+4f_\mathbb{T}\mathbb{L}_m
-2f_\mathbb{T}g^{\rho\beta}g^{\zeta\eta}\frac{\partial^2\mathbb{L}_m}{\partial
g^{\rho\beta}\partial g^{\zeta\eta}}=0.
\end{align}
The field equations and the corresponding results can be achieved in
$f(\mathbb{R})$ gravity by considering the vacuum case. The
curvature-matter coupling in this extended theory produces non-null
divergence of the $\mathbb{EMT}$ due to which there exists an extra
force in the gravitational field, and thus opposing $\mathbb{GR}$
and $f(\mathbb{R})$ gravity. Consequently, we obtain
\begin{align}\nonumber
\nabla^\zeta\mathbb{T}_{\zeta\eta}&=\frac{f_\mathbb{T}}{8\pi-f_\mathbb{T}}\bigg[(\mathbb{T}_{\zeta\eta}
+\Theta_{\zeta\eta})\nabla^\zeta\ln{f_\mathbb{T}}+\nabla^\zeta\Theta_{\zeta\eta}\\\label{g11}
&-\frac{8\pi\alpha}{f_\mathbb{T}}\nabla^\zeta\mathfrak{A}_{\zeta\eta}-\frac{1}{2}g_{\rho\beta}\nabla_\eta\mathbb{T}^{\rho\beta}\bigg],
\end{align}
where
$\Theta_{\zeta\eta}=g_{\zeta\eta}\mathbb{L}_m-2\mathbb{T}_{\zeta\eta}-2g^{\rho\beta}\frac{\partial^2
\mathbb{L}_{m}}{\partial g^{\zeta\eta}\partial g^{\rho\beta}}$ and
we consider $\mathbb{L}_{m}=P=\frac{P_r+2P_\bot}{3}$ in this case
which leads to $\frac{\partial^2 \mathbb{L}_{m}}{\partial
g^{\zeta\eta}\partial g^{\rho\beta}}=0$.

The hypersurface $\Sigma$ distinguishes the inner and outer regions
of a geometrical structure, thus the metric defining the interior
spherical configuration is given as
\begin{equation}\label{g6}
ds^2=-e^{\sigma} dt^2+e^{\chi} dr^2+r^2d\theta^2+r^2\sin^2\theta
d\vartheta^2,
\end{equation}
where $\sigma=\sigma(r)$ and $\chi=\chi(r)$. This line element
determines the corresponding four-vector and four-velocity as
\begin{equation}\label{g7}
\mathcal{W}^\zeta=(0,e^{\frac{-\chi}{2}},0,0), \quad
\mathcal{K}^\zeta=(e^{\frac{-\sigma}{2}},0,0,0),
\end{equation}
satisfying the relations $\mathcal{W}^\zeta
\mathcal{K}_{\zeta}=0,~\mathcal{W}^\zeta \mathcal{W}_{\zeta}=1$ and
$\mathcal{K}^\zeta \mathcal{K}_{\zeta}=-1$. We adopt a particular
model of $f(\mathbb{R},\mathbb{T})$ gravity to express our results
in a meaningful way. The linear model in this regard provides the
entire structural transformation of self-gravitating objects, thus
we consider
\begin{equation}\label{g61}
f(\mathbb{R},\mathbb{T})=f_1(\mathbb{R})+
f_2(\mathbb{T})=\mathbb{R}+2\varpi\mathbb{T},
\end{equation}
where $\varpi$ is an arbitrary coupling constant and
$\mathbb{T}=-\mu+P_r+2P_\bot$. Houndjo and Piattella \cite{38}
analyzed the pressureless matter configuration and found that the
characteristics of holographic dark energy may be reproduced through
this model. Moraes et al. \cite{39} observed this model compatible
with standard conservation of the $\mathbb{EMT}$.

The presence of additional source makes the field equations
corresponding to metric \eqref{g6} and gravitational model
\eqref{g61} as
\begin{align}\label{g8}
&e^{-\chi}\left(\frac{\chi'}{r}-\frac{1}{r^2}\right)
+\frac{1}{r^2}=8\pi\left(\mu-\alpha\mathfrak{A}_{0}^{0}\right)+\varpi\left(3\mu-\frac{P_r}{3}-\frac{2P_\bot}{3}\right),\\\label{g9}
&e^{-\chi}\left(\frac{1}{r^2}+\frac{\sigma'}{r}\right)
-\frac{1}{r^2}=8\pi\left(P_r+\alpha\mathfrak{A}_{1}^{1}\right)-\varpi\left(\mu-\frac{7P_r}{3}-\frac{2P_\bot}{3}\right),
\\\label{g10}
&\frac{e^{-\chi}}{4}\left[\sigma'^2-\chi'\sigma'+2\sigma''-\frac{2\chi'}{r}+\frac{2\sigma'}{r}\right]
=8\pi\left(P_\bot+\alpha\mathfrak{A}_{2}^{2}\right)-\varpi\left(\mu-\frac{P_r}{3}-\frac{8P_\bot}{3}\right),
\end{align}
where the last terms on right hand side of the above equations
represent the $f(\mathbb{R},\mathbb{T})$ corrections and prime means
$\frac{\partial}{\partial r}$. Moreover, for the model \eqref{g61},
Eq.\eqref{g11} yields
\begin{align}\nonumber
&\frac{dP_r}{dr}+\frac{\sigma'}{2}\left(\mu+P_r\right)+\frac{\alpha\sigma'}{2}
\left(\mathfrak{A}_{1}^{1}-\mathfrak{A}_{0}^{0}\right)+\frac{2}{r}\left(P_r-P_\bot\right)\\\label{g12}
&+\alpha\frac{d\mathfrak{A}_{1}^{1}}{dr}+\frac{2\alpha}{r}\left(\mathfrak{A}_{1}^{1}
-\mathfrak{A}_{2}^{2}\right)=-\frac{\varpi}{4\pi-\varpi}\big(\mu'-P'\big),
\end{align}
which confirms the non-conserved nature of this extended theory.
Equation \eqref{g12} helps in studying structural changes of
self-gravitating object and also named as the generalization of
Tolman-Opphenheimer-Volkoff equation. The set of field equations
become complicated due to the inclusion of new source as the number
of unknowns increase, i.e.,
$(\sigma,\chi,\mu,P_r,P_\bot,\mathfrak{A}_{0}^{0},\mathfrak{A}_{1}^{1},\mathfrak{A}_{2}^{2})$,
thus this system cannot be solved analytically unless we use some
constraints. We employe a systematic approach \cite{33} in this
regard to make the field equations solvable.

\section{Gravitational Decoupling}

An effective approach, known as the gravitational decoupling, allows
the transformation of the metric potentials and helps in obtaining
the solution of the considered matter source. To implement this
technique, we consider a solution to Eqs.\eqref{g8}-\eqref{g10} by
the following metric
\begin{equation}\label{g15}
ds^2=-e^{\rho(r)}dt^2+\frac{1}{\xi(r)}dr^2+r^2d\theta^2+r^2\sin^2\theta
d\vartheta^2.
\end{equation}
The linear form of decoupling transformations are
\begin{equation}\label{g16}
\rho\rightarrow\sigma=\rho+\alpha\mathrm{t}, \quad \xi\rightarrow
e^{-\chi}=\xi+\alpha\mathrm{f},
\end{equation}
where $\mathrm{f}$ and $\mathrm{t}$ deform the radial and temporal
component, respectively. We employ the MGD scheme in the current
setup, thus only the radial metric coefficient is allowed to
transform whereas the temporal one remains preserved, i.e.,
$\mathrm{t}\rightarrow 0,~\mathrm{f}\rightarrow \mathcal{F}$.
Equation \eqref{g16} then reduces to
\begin{equation}\label{g17}
\rho\rightarrow\sigma=\rho, \quad \xi\rightarrow
e^{-\chi}=\xi+\alpha \mathcal{F},
\end{equation}
where $\mathcal{F}=\mathcal{F}(r)$. It is mentioned here that the
spherical symmetry is not disturbed by such linear mapping. After
applying the transformation \eqref{g17} in the field equations
\eqref{g8}-\eqref{g10}, the first set (corresponding to $\alpha=0$)
representing the anisotropic seed source is obtained as
\begin{align}\label{g18}
&e^{-\chi}\left(\frac{\chi'}{r}-\frac{1}{r^2}\right)
+\frac{1}{r^2}=8\pi\mu+\varpi\left(3\mu-\frac{P_r}{3}-\frac{2P_\bot}{3}\right),\\\label{g19}
&e^{-\chi}\left(\frac{1}{r^2}+\frac{\sigma'}{r}\right)
-\frac{1}{r^2}=8\pi
P_r-\varpi\left(\mu-\frac{7P_r}{3}-\frac{2P_\bot}{3}\right),
\\\label{g20}
&\frac{e^{-\chi}}{4}\left[\sigma'^2-\chi'\sigma'+2\sigma''-\frac{2\chi'}{r}+\frac{2\sigma'}{r}\right]
=8\pi P_\bot-\varpi\left(\mu-\frac{P_r}{3}-\frac{8P_\bot}{3}\right),
\end{align}
from which the state variables can explicitly be extracted as
\begin{align}\nonumber
\mu&=\frac{e^{-\chi}}{48r^2(\varpi+2\pi)(\varpi+4\pi)}\big[2\big\{r^2\varpi\sigma''+8r\chi'(\varpi+3\pi)+8(\varpi+3\pi)\\\label{g18a}
&\times\big(e^{\chi}-1\big)\big\}+r^2\varpi\sigma'^2+r\varpi\sigma'\big(4-r\chi'\big)\big],\\\nonumber
P_r&=\frac{e^{-\chi}}{48r^2(\varpi+2\pi)(\varpi+4\pi)}\big[2\big\{4r\varpi\chi'-r^2\varpi\sigma''
-8(\varpi+3\pi)\big(e^{\chi}-1\big)\big\}\\\label{g19a}
&-r^2\varpi\sigma'^2+r\sigma'\big(20\varpi+r\varpi\chi'+48\pi\big)\big],\\\nonumber
P_\bot&=\frac{e^{-\chi}}{48r^2(\varpi+2\pi)(\varpi+4\pi)}\big[10r^2\varpi\sigma''+(5\varpi+12\pi)r^2\sigma'^2+24\pi{r^2}\sigma''\\\label{g20a}
&-8\varpi+r\sigma'\big\{8(\varpi+3\pi)-(5\varpi+12\pi)r\chi'\big\}-4r\chi'(\varpi+6\pi)+8{\varpi}e^{\chi}\big].
\end{align}

On the other hand, the influence of additional source
($\mathfrak{A}^{\zeta}_{\eta}$) is encoded in the following set (for
$\alpha=1$) as
\begin{align}\label{g21}
&8\pi\mathfrak{A}_{0}^{0}=\frac{\mathcal{F}'}{r}+\frac{\mathcal{F}}{r^2},\\\label{g22}
&8\pi\mathfrak{A}_{1}^{1}=\mathcal{F}\left(\frac{\sigma'}{r}+\frac{1}{r^2}\right),\\\label{g23}
&8\pi\mathfrak{A}_{2}^{2}=\frac{\mathcal{F}}{4}\left(2\sigma''+\sigma'^2+\frac{2\sigma'}{r}\right)
+\mathcal{F}'\left(\frac{\sigma'}{4}+\frac{1}{2r}\right).
\end{align}
The MGD scheme does not allow the exchange of energy between the two
(original and additional) matter sources and they are conserved
individually. The system \eqref{g8}-\eqref{g10} has successfully
been decoupled into two sets. The first set
\eqref{g18a}-\eqref{g20a} contains five unknowns
($\mu,P_r,P_\bot,\sigma,\chi$), thus a well-behaved solution will be
assumed to close it. The second sector \eqref{g21}-\eqref{g23}
involves four unknowns
($\mathcal{F},\mathfrak{A}_{0}^{0},\mathfrak{A}_{1}^{1},\mathfrak{A}_{2}^{2}$),
so that a constraint on $\mathfrak{A}$-sector will be helpful to
reduce the number of undetermined quantities. The effective matter
variable can clearly be identified as
\begin{equation}\label{g13}
\tilde{\mu}=\mu-\alpha\mathfrak{A}_{0}^{0},\quad
\tilde{P}_{r}=P_r+\alpha\mathfrak{A}_{1}^{1}, \quad
\tilde{P}_{\bot}=P_\bot+\alpha\mathfrak{A}_{2}^{2},
\end{equation}
which lead to the total anisotropy of the system as
\begin{equation}\label{g14}
\tilde{\Pi}=\tilde{P}_{\bot}-\tilde{P}_{r}=(P_\bot-P_r)+\alpha(\mathfrak{A}_{2}^{2}-\mathfrak{A}_{1}^{1})=\Pi+\Pi_{\mathfrak{A}},
\end{equation}
where the seed and new sources generate the anisotropy $\Pi$ and
$\Pi_{\mathfrak{A}}$, respectively.

\section{Isotropization of Compact Sources}

It can be noticed from Eq.\eqref{g14} that $\tilde{\Pi}$ is the
total anisotropy generated by the system which may differ from that
of generated by the seed source $\mathbb{T}_{\zeta\eta}$, i.e.,
$\Pi$. In this section, we consider that the anisotropic structure
converts into an isotropic one ($\tilde{\Pi}=0$) after the inclusion
of new source. The variation in parameter $\alpha$ controls this
change, as $\alpha=0$ and $1$ represent the anisotropic and
isotropic structures, respectively. Here, we discuss the case when
$\alpha=1$ which yields
\begin{equation}\label{g14a}
\Pi_{\mathfrak{A}}=-\Pi \quad \Rightarrow \quad
\mathfrak{A}_{2}^{2}-\mathfrak{A}_{1}^{1}=P_r-P_\bot.
\end{equation}
Casadio et al. \cite{37k} recently used this condition to isotropize
the system that was initially considered to be anisotropic with the
help of gravitational decoupling. In order to get the first
solution, we take a particular ansatz related to the seed
anisotropic source as
\begin{eqnarray}\label{g33}
\sigma(r)&=&\ln\bigg\{\mathcal{B}^2\bigg(1+\frac{r^2}{\mathcal{A}^2}\bigg)\bigg\},\\\label{g34}
\xi(r)&=&e^{-\chi}=\frac{\mathcal{A}^2+r^2}{\mathcal{A}^2+3r^2},\\\label{g35}
\mu&=&\frac{3\big(3\varpi+8\pi\big)\mathcal{A}^2+2\big(5\varpi+12\pi\big)r^2}{4\big(\varpi+2\pi\big)
\big(\varpi+4\pi\big)\big(\mathcal{A}^2+3r^2\big)^2},\\\label{g36}
P_r&=&\frac{\varpi\big(3\mathcal{A}^2+2r^2\big)}{4\big(\varpi+2\pi\big)\big(\varpi+4\pi\big)\big(\mathcal{A}^2+3r^2\big)^2},\\\label{g37}
P_\bot&=&\frac{3\varpi\mathcal{A}^2+8{\varpi}r^2+12{\pi}r^2}{4\big(\varpi+2\pi\big)\big(\varpi+4\pi\big)\big(\mathcal{A}^2+3r^2\big)^2},
\end{eqnarray}
where $\mathcal{A}^2$ and $\mathcal{B}^2$ are undetermined constants
and we calculate them through smooth matching. The gravitational
field, in which the cluster of particles move in arbitrarily
oriented circular orbits, can be determined by means of the above
solution \cite{42a1}. This spacetime has also been used to construct
decoupled solutions in the context of $\mathbb{GR}$ \cite{37k}.

The junction conditions play significant role in examining different
characteristics of stellar bodies at the hypersurface
($\Sigma:r=\mathcal{R}$). Thus, we consider the Schwarzschild
exterior spacetime to match the inner and outer regions smoothly
\begin{equation}\label{g25}
ds^2=-\frac{r-2\tilde{\mathcal{M}}}{r}dt^2+\frac{r}{r-2\tilde{\mathcal{M}}}dr^2+
r^2d\theta^2+r^2\sin^2\theta d\vartheta^2,
\end{equation}
where $\tilde{\mathcal{M}}$ indicates the total mass. We obtain two
unknowns through matching conditions as
\begin{eqnarray}\label{g37}
\mathcal{A}^2&=&\frac{\mathcal{R}^2\big(\mathcal{R}-3\tilde{\mathcal{M}}\big)}{\tilde{\mathcal{M}}},\\\label{g38}
\mathcal{B}^2&=&\frac{\mathcal{R}-2\tilde{\mathcal{M}}}{\frac{\tilde{\mathcal{M}}\mathcal{R}}{\mathcal{R}-3\tilde{\mathcal{M}}}+\mathcal{R}}.
\end{eqnarray}
Further, we analyze the physical feasibility of a particular compact
star, namely $4U 1820-30$ having mass $\tilde{\mathcal{M}}=1.58 \pm
0.06 M_{\bigodot}$ and radius $\mathcal{R}=9.1 \pm 0.4 km$
\cite{42aa}. All the graphical observations are done by using this
data. The condition \eqref{g14a} along with the field equations and
metric functions \eqref{g33}-\eqref{g34} provides the differential
equation as
\begin{align}\nonumber
&r\big(\mathcal{A}^2+r^2\big)\big\{(\varpi+4\pi)\big(\mathcal{A}^2+2r^2\big)\big(\mathcal{A}^2+3r^2\big)^2\mathcal{F}'(r)+24\pi
r^3 \big(\mathcal{A}^2+r^2\big)\big\}\\\label{g39}
&-2(\varpi+4\pi)\big(\mathcal{A}^2+3r^2\big)^2\big(\mathcal{A}^4+2\mathcal{A}^2r^2+2r^4\big)\mathcal{F}(r)=0,
\end{align}
whose analytical solution is
\begin{equation}\label{g40}
\mathcal{F}(r)=\frac{r^2\big(\mathcal{A}^2+r^2\big)}{\mathcal{A}^2+2r^2}\bigg\{\mathbb{C}_1+\frac{4\pi}{(\varpi+4\pi)
\big(\mathcal{A}^2+3r^2\big)}\bigg\},
\end{equation}
where $\mathbb{C}_1$ is treated as the integration constant. The
deformed radial metric component \eqref{g17} takes the form
\begin{align}\nonumber
e^{\chi}&=\xi^{-1}=\big[\big(\mathcal{A}^2+r^2\big)\big\{\alpha{r^2}\big(\varpi\mathbb{C}_{1}\big(\mathcal{A}^2+3r^2\big)
+4\pi\big(\mathcal{A}^2\mathbb{C}_{1}+3\mathbb{C}_{1}r^2+1\big)\big)\\\label{g40a}
&+(\varpi+4\pi)\big(\mathcal{A}^2+2r^2\big)\big\}\big]^{-1}\big[\big(\varpi+4\pi\big)\big(\mathcal{A}^2+2r^2\big)
\big(\mathcal{A}^2+3r^2\big)\big].
\end{align}

Hence, the minimally deformed solution to the system
\eqref{g8}-\eqref{g10} can be represented by the spacetime given as
\begin{equation}\label{g41}
ds^2=-\mathcal{B}^2\bigg(1+\frac{r^2}{\mathcal{A}^2}\bigg)dt^2+\frac{\mathcal{A}^2+3r^2}
{\mathcal{A}^2+r^2+\alpha\mathcal{F}\big(\mathcal{A}^2+3r^2\big)}dr^2+
r^2d\theta^2+r^2\sin^2\theta d\vartheta^2,
\end{equation}
whose state variables (such as the energy density and pressure
components) are
\begin{eqnarray}\nonumber
\tilde{\mu}&=&\frac{-1}{8\pi\big(\varpi+2\pi\big)\big(\varpi+4\pi\big)\big(\mathcal{A}^2+2r^2\big)^2
\big(\mathcal{A}^2+3r^2\big)^2}\big[\big(\varpi^2+6\pi\varpi+8\pi^2\big)\\\nonumber
&\times&3\mathcal{A}^8\alpha\mathbb{C}_{1}+\mathcal{A}^6 \big\{6 \pi
\varpi \big(2 \alpha +25 \alpha \mathbb{C}_{1} r^2-3\big)+8 \pi
^2\big(3 \alpha +25 \alpha \mathbb{C}_{1} r^2-6\big)\\\nonumber
&+&25 \varpi ^2 \alpha  C_{1} r^2 \big\}+\mathcal{A}^4 r^2 \big\{2
\pi \varpi \big(5 \alpha \big(45 \mathbb{C}_{1}
r^2+4\big)-46\big)+75 \varpi ^2 \alpha \mathbb{C}_{1} r^2\\\nonumber
&+&40 \pi ^2 \big(\alpha \big(15 \mathbb{C}_{1}
r^2+2\big)-6\big)\big\}+\mathcal{A}^2 r^4 \big\{2 \pi \varpi \big(9
\alpha \big(33 \mathbb{C}_{1} r^2+2\big)-76\big)\\\nonumber &+&99
\varpi ^2 \alpha \mathbb{C}_{1} r^2+24 \pi ^2 \big(\alpha \big(33
\mathbb{C}_{1} r^2+3\big)-16\big)\big\}+2 r^6 \big\{27 \varpi ^2
\alpha \mathbb{C}_{1} r^2\\\label{g46} &+&2 \pi \varpi \big(6 \alpha
+81 \alpha \mathbb{C}_{1} r^2-20\big)+24 \pi ^2 \big(\alpha +9
\alpha \mathbb{C}_{1} r^2-4\big)\big\}\big],\\\nonumber
\tilde{P}_{r}&=&\frac{1}{8\pi\big(\varpi+2\pi\big)\big(\varpi+4\pi\big)\big(\mathcal{A}^2+2r^2\big)
\big(\mathcal{A}^2+3r^2\big)^2}\big[ \alpha\big(\mathcal{A}^2+3
r^2\big)^2\\\nonumber &\times& \big(\varpi +2 \pi \big) \big\{\varpi
\mathbb{C}_{1} \big(\mathcal{A}^2+3 r^2\big)+4 \pi
\big(\mathcal{A}^2 \mathbb{C}_{1}+3 \mathbb{C}_{1}
r^2+1\big)\big\}+2 \pi \varpi\\\label{g47} &\times&
\big(\mathcal{A}^2+2r^2\big)\big(3\mathcal{A}^2+2r^2\big)\big],
\\\nonumber \tilde{P}_{\bot}&=&\frac{1}{8\pi\big(\varpi+2\pi\big)\big(\varpi+4\pi\big)\big(\mathcal{A}^2+2r^2\big)
\big(\mathcal{A}^2+3r^2\big)^2}\big[\big(\varpi ^2+6 \pi  \varpi +8
\pi ^2\big)\\\nonumber &\times&\mathcal{A}^6 \alpha
\mathbb{C}_{1}+\mathcal{A}^4 \big\{2 \pi \varpi \big(2 \alpha +27
\alpha\mathbb{C}_{1} r^2+3\big)+8 \pi ^2 \big(\alpha +9
\alpha\mathbb{C}_{1} r^2\big)\\\nonumber &+&9 \varpi ^2
\alpha\mathbb{C}_{1} r^2\big\}+\mathcal{A}^2 r^2 \big\{27 \varpi ^2
\alpha\mathbb{C}_{1} r^2+2 \pi \varpi \big(6 \alpha +81
\alpha\mathbb{C}_{1} r^2+14\big)\\\nonumber &+&24 \pi ^2 \big(\alpha
+9 \alpha \mathbb{C}_{1} r^2+1\big)\big\}+r^4 \big\{27 \varpi ^2
\alpha\mathbb{C}_{1} r^2+24 \pi ^2 \big(\alpha +9 \alpha
\mathbb{C}_{1} r^2+2\big)\\\label{g48} &+&2 \pi \varpi \big(6 \alpha
+81 \alpha\mathbb{C}_{1} r^2+16\big)\big\}\big],
\end{eqnarray}
and the corresponding anisotropy is
\begin{eqnarray}\label{g49}
\tilde{\Pi}&=&\frac{3r^2\big(1-\alpha\big)}{2\big(\varpi+4\pi\big)
\big(\mathcal{A}^2+3r^2\big)^2},
\end{eqnarray}
which disappears for $\alpha=1$. Equations \eqref{g46}-\eqref{g49}
provide exact solution of the $f(\mathbb{R},\mathbb{T})$ field
equations for $\alpha\in[0,1]$. It can be observed that the system
is initially anisotropic at $\alpha=0$ which is then deformed into
the isotropic one ($\alpha=1$). Hence, the process of isotropization
can be followed in detail by varying this parameter between $0$ and
$1$.

\section{Complexity of Compact Sources}

The definition of complexity for static spherical structure
\cite{37g} has also been extended to the dynamical scenario
\cite{37h}. The key feature of this notion is that the
uniform/isotropic configuration is assigned a zero value of the
complexity factor. The orthogonal splitting of the curvature tensor
results in certain scalars, from which $\mathbb{Y}_{TF}$ is found to
be the complexity factor for self-gravitating spacetime. This factor
in terms of the inhomogeneous energy density and pressure anisotropy
along with $f(\mathbb{R},\mathbb{T})$ terms has the form
\begin{equation}\label{g51}
\mathbb{Y}_{TF}(r)=8\pi\Pi\big(1+\varpi\big)-\frac{4\pi}{r^3}\int_0^rz^3\mu'(z)dz.
\end{equation}
The Tolman mass is generally defined as
\begin{equation}\label{g52}
m_T=4\pi\int_0^rz^2e^{\frac{(\sigma+\chi)}{2}}\big(\mu+P_r+2P_\bot\big)dz,
\end{equation}
which gives the total energy of the fluid contained in sphere. The
Tolman mass can be written together with the complexity factor as
\begin{equation}\label{g53}
m_T=\tilde{\mathcal{M}}_T\left(\frac{r}{\mathcal{R}}\right)^3+r^3\int_r^{\mathcal{R}}\frac{e^{\frac{(\sigma+\chi)}{2}}}{z}
\big(\mathbb{Y}_{TF}-8\pi\Pi\varpi\big)dz,
\end{equation}
where $\tilde{\mathcal{M}}_T$ is the total Tolman mass.

The complexity factor for the considered source
\eqref{g8}-\eqref{g10} becomes
\begin{eqnarray}\nonumber
\tilde{\mathbb{Y}}_{TF}(r)&=&8\pi\tilde{\Pi}\big(1+\varpi\big)-\frac{4\pi}{r^3}\int_0^rz^3\tilde{\mu}'(z)dz\\\nonumber
&=&8\pi\Pi\big(1+\varpi\big)-\frac{4\pi}{r^3}\int_0^rz^3\mu'(z)dz\\\label{g54}
&+&8\pi\Pi_{\mathfrak{A}}\big(1+\varpi\big)+\frac{4\pi}{r^3}\int_0^rz^3\mathfrak{A}{_0^0}'(z)dz,
\end{eqnarray}
which can equivalently be written as
\begin{eqnarray}\label{g55}
\tilde{\mathbb{Y}}_{TF}=\mathbb{Y}_{TF}+\mathbb{Y}_{TF}^{\mathfrak{A}},
\end{eqnarray}
where $\mathbb{Y}_{TF}$ and $\mathbb{Y}_{TF}^{\mathfrak{A}}$
correspond to the systems \eqref{g18a}-\eqref{g20a} and
\eqref{g21}-\eqref{g23}, respectively. Since we establish the
solution \eqref{g46}-\eqref{g49} for $\tilde{\Pi}=0$, so that
Eq.\eqref{g54} yields
\begin{eqnarray}\label{g56}
\tilde{\mathbb{Y}}_{TF}&=&-\frac{4\pi}{r^3}\int_0^rz^3\tilde{\mu}'(z)dz,
\end{eqnarray}
which gives the complexity factor for the metric \eqref{g41} after
combining with the field equations as
\begin{eqnarray}\nonumber
\tilde{\mathbb{Y}}_{TF}&=&-\frac{1}{6r^3\big(\varpi+2\pi\big)\big(\varpi+4\pi\big)\big(\mathcal{A}^2+2r^2\big)^2
\big(\mathcal{A}^2+3r^2\big)^2}\bigg[\pi\varpi\sqrt{3\mathcal{A}^2}\\\nonumber
&\times&\big(\mathcal{A}^4+5\mathcal{A}^2r^2+6r^4\big)^2\tan^{-1}\bigg(\frac{\sqrt{3}r}{\sqrt{\mathcal{A}}^2}\bigg)-3
\big\{\pi\varpi\mathcal{A}^8r-\mathcal{A}^6\big(2\varpi^2\alpha\mathbb{C}_{1}r^5\\\nonumber
&+&16\pi^2\alpha\mathbb{C}_{1}r^5+3\pi\varpi{r^3}\big(4\alpha\mathbb{C}_{1}r^2-3\big)\big)
-4\mathcal{A}^4r^5\big(3\varpi^2\alpha\mathbb{C}_{1}r^2+2\pi\varpi\\\nonumber
&\times&\big(4\alpha+9\alpha\mathbb{C}_{1}r^2-8\big)+8\pi^2\big(2\alpha+3\alpha\mathbb{C}_{1}r^2-3\big)\big)
-6\mathcal{A}^2r^7\big(3\varpi^2\alpha\mathbb{C}_{1}r^2\\\nonumber
&+&2\pi\varpi\big(8\alpha+9\alpha\mathbb{C}_{1}r^2-15\big)+8\pi^2\big(4\alpha+3\alpha\mathbb{C}_{1}r^2-8\big)\big)+16\pi{r^9}\\\label{g56a}
&\times&\big(\varpi\big(10-3\alpha\big)-6\pi\big(\alpha-4))\big\}\bigg].
\end{eqnarray}

\subsection{Two Systems with the Same Complexity Factor}

Here, we assume that the complexity factor $\mathbb{Y}_{TF}$
associated with the seed source remains unchanged after the addition
of new source $\mathfrak{A}^{\zeta}_{\eta}$, i.e.,
$\mathbb{Y}_{TF}^{\mathfrak{A}}=0$ which leads to
$\tilde{\mathbb{Y}}_{TF}=\mathbb{Y}_{TF}$ or
\begin{eqnarray}\label{g56b}
8\pi\Pi_{\mathfrak{A}}\big(1+\varpi\big)=-\frac{4\pi}{r^3}\int_0^rz^3\mathfrak{A}{_0^0}'(z)dz.
\end{eqnarray}
The right side of Eq.\eqref{g56b} along with \eqref{g21} becomes
\begin{eqnarray}\label{g56c}
-\frac{4\pi}{r^3}\int_0^rz^3\mathfrak{A}{_0^0}'(z)dz=\frac{\mathcal{F}}{r^2}-\frac{\mathcal{F}'}{2r},
\end{eqnarray}
and the condition \eqref{g56b} results in the differential equation
of first order as
\begin{eqnarray}\nonumber
&&(\varpi+1)\bigg\{\mathcal{F}'(r)\bigg(\frac{\sigma'}{4}+\frac{1}{2r}\bigg)+\mathcal{F}(r)\bigg(\frac{\sigma''}{2}-\frac{1}{r^2}
+\frac{\sigma'^2}{4}-\frac{\sigma'}{2r}\bigg)\bigg\}\\\label{g56d}
&&+\frac{1}{2}\bigg(\frac{\mathcal{F}'(r)}{r}-\frac{2\mathcal{F}(r)}{r^2}\bigg)=0.
\end{eqnarray}
We observe from this equation that its solution depends on the
metric function that describes the seed source
$\mathbb{T}_{\zeta\eta}$. Thus, we use the Tolman IV ansatz to
formulate the corresponding solution as
\begin{align}\label{g57}
\sigma(r)&=\ln\bigg\{\mathcal{B}^2\bigg(1+\frac{r^2}{\mathcal{A}^2}\bigg)\bigg\},\\\label{g58}
\xi(r)&=e^{-\chi}=\frac{\big(\mathcal{A}^2+r^2\big)\big(\mathcal{C}^2-r^2\big)}{\mathcal{C}^2\big(\mathcal{A}^2+2r^2\big)},
\end{align}
generated by the energy density and isotropic pressure
\begin{align}\nonumber
\mu&=\frac{1}{4\mathcal{C}^2\big(\varpi+2\pi\big)\big(\varpi+4\pi\big)\big(\mathcal{A}^2+2r^2\big)^2}
\big[4\big(\varpi+3\pi\big)\mathcal{A}^4+\mathcal{A}^2\\\nonumber
&\times\big(8\varpi{r^2}+28\pi{r^2}+5\varpi\mathcal{C}^2+12\pi\mathcal{C}^2\big)+2r^2\big\{\varpi\big(3r^2+2\mathcal{C}^2\big)\\\label{g59}
&+4\pi\big(3r^2+\mathcal{C}^2\big)\big\}\big],\\\nonumber
P&=\frac{1}{4\mathcal{C}^2\big(\varpi+2\pi\big)\big(\varpi+4\pi\big)\big(\mathcal{A}^2+2r^2\big)^2}
\big[\mathcal{A}^2\big\{4\pi\big(\mathcal{C}^2-5r^2\big)-4\varpi{r^2}\\\label{g60}
&+3\varpi\mathcal{C}^2\big\}-4\pi\mathcal{A}^4+2r^2\big\{2\varpi\mathcal{C}^2+4\pi\big(\mathcal{C}^2-3r^2\big)-3\varpi{r^2}\big\}\big].
\end{align}
The constants $\mathcal{A}^2$ and $\mathcal{B}^2$ are the same as
provided in Eqs.\eqref{g37} and \eqref{g38}, while $\mathcal{C}^2$
is obtained as
\begin{align}\label{g60a}
\mathcal{C}^2&=\frac{\mathcal{R}^3}{\tilde{\mathcal{M}}}.
\end{align}
After plugging the metric potential \eqref{g57} in differential
equation \eqref{g56d}, we have
\begin{equation}\label{g60b}
\mathcal{F}(r)=\frac{\mathbb{C}_2r^2\big(\mathcal{A}^2+r^2\big)}{\mathcal{A}^2\big(2+\varpi\big)+r^2\big(2\varpi+3\big)},
\end{equation}
where $\mathbb{C}_2$ is the integration constant. Hence, the
deformed radial metric component takes the form
\begin{align}\label{g60c}
e^{\chi}&=\xi^{-1}=\frac{\mathcal{C}^2\big(\mathcal{A}^2+2r^2\big)}{\big(\mathcal{A}^2+r^2\big)}\bigg\{\mathcal{C}^2
+r^2\bigg(\frac{\alpha\mathbb{C}_2\mathcal{C}^2\big(\mathcal{A}^2+2r^2\big)}{\mathcal{A}^2\big(2+\varpi\big)+r^2\big(2\varpi+3\big)}
-1\bigg)\bigg\}^{-1}.
\end{align}
The definition of $\tilde{\mathbb{Y}}_{TF}$ for the considered setup
is given in Eq.\eqref{g54} which leads to
\begin{align}\nonumber
\tilde{\mathbb{Y}}_{TF}&=\mathbb{Y}_{TF}=\frac{\pi\big(\mathcal{A}^2+2\mathcal{C}^2\big)}{16r^3\mathcal{C}^2\big(\varpi+2\pi\big)
\big(\varpi+4\pi\big)\big(\mathcal{A}^2+2r^2\big)^2}\bigg[6\varpi\mathcal{A}^4r+20\varpi\mathcal{A}^2r^3\\\label{g60d}
&-3\varpi\sqrt{2\mathcal{A}^2}\big(\mathcal{A}^2+2r^2\big)^2\tan^{-1}\bigg(\frac{\sqrt{2}r}{\sqrt{\mathcal{A}^2}}\bigg)
+64\varpi{r^5}+128\pi r^5\bigg].
\end{align}

\subsection{Generating Solutions with Zero Complexity}

In this section, we formulate a solution to the modified field
equations corresponding to $\tilde{\mathbb{Y}}_{TF}=0$. We consider
that the seed source is not complexity-free, i.e.,
$\mathbb{Y}_{TF}\neq0$, but the addition of new source results in
the vanishing complexity factor. Hence, the total matter
configuration has zero complexity due to which Eq.\eqref{g55} in
terms of Tolman IV ansatz gives rise to
\begin{align}\nonumber
&\frac{8r\big(\varpi+1\big)}{\big(\mathcal{A}^2+r^2\big)^2}\big[r\big(\mathcal{A}^4+3\mathcal{A}^2r^2+2r^4\big)\mathcal{F}'(r)-2
\big(\mathcal{A}^4+2\mathcal{A}^2r^2+2r^4\big)\mathcal{F}(r)\big]\\\nonumber
&+\frac{\pi}{\big(\varpi+2\pi\big)\big(\varpi+4\pi\big)\mathcal{C}^2\big(\mathcal{A}^2+2r^2\big)^2}
\bigg[\big(\mathcal{A}^2+2\mathcal{C}^2\big)\big\{6\varpi\mathcal{A}^4r+20\varpi\mathcal{A}^2r^3\\\nonumber
&-3\varpi\sqrt{2\mathcal{A}^2}\big(\mathcal{A}^2+2r^2\big)^2\tan^{-1}\bigg(\frac{\sqrt{2}r}{\sqrt{\mathcal{A}^2}}\bigg)
+64\varpi{r^5}+128\pi{r^5}\big\}\bigg]\\\label{60e}
&+8r\big(r\mathcal{F}'(r)-2\mathcal{F}(r)\big)=0,
\end{align}
whose solution is given by
\begin{align}\nonumber
\mathcal{F}(r)&=-\frac{\pi{r^2}\big(\mathcal{A}^2+r^2\big)\big(\mathcal{A}^2+2\mathcal{C}^2\big)
\bigg\{\frac{\varpi\sqrt{2\mathcal{A}^2}\tan^{-1}\big(\frac{\sqrt{2}
r}{\sqrt{A}}\big)}{r^3}-\frac{4(3\varpi+8\pi)}{\mathcal{A}^2+2r^2}-\frac{2\varpi}{r^2}\bigg\}}{8\mathcal{C}^2
\big(\varpi^2+6\pi\varpi+8\pi^2\big)\big(\varpi\mathcal{A}^2+2\mathcal{A}^2+2\varpi{r^2}+3r^2\big)}\\\label{60f}
&+\frac{\mathbb{C}_3r^2\big(\mathcal{A}^2+r^2\big)}{\varpi\mathcal{A}^2+2\mathcal{A}^2+2\varpi
r^2+3r^2},
\end{align}
where $\mathbb{C}_3$ indicates the integration constant with
dimension of inverse square length. The radial coefficient can be
deformed by using the transformation \eqref{g17} along with the
above equation as
\begin{align}\label{g60fa}
e^{\chi}&=\xi^{-1}=\frac{\mathcal{C}^2\big(\mathcal{A}^2+2r^2\big)}{\big(\mathcal{A}^2+r^2\big)\big(\mathcal{C}^2-r^2\big)
+\alpha\mathcal{C}^2\mathcal{F}(r)\big(\mathcal{A}^2+2r^2\big)}.
\end{align}
The final form of the matter variables corresponding to the
deformation function \eqref{60f} is
\begin{align}\nonumber
\tilde{\mu}&=\frac{1}{32\mathcal{C}^2\big(\mathcal{A}^2+2r^2\big)^2}\bigg[\frac{8}{\big(\varpi+2\pi\big)\big(\varpi+4\pi\big)}
\big\{4\big(\varpi+3\pi\big)\mathcal{A}^4+\mathcal{A}^2\big(8\varpi{r^2}\\\nonumber
&+28\pi{r^2}+5\varpi\mathcal{C}^2+12\pi\mathcal{C}^2\big)+2r^2\big(\varpi
\big(3r^2+2\mathcal{C}^2\big)+4\pi\big(3r^2+\mathcal{C}^2\big)\big)\big\}\\\nonumber
&-\frac{\alpha}{\pi\big(\varpi^2+6\pi\varpi+8\pi^2\big)r\big(\big(\varpi+2\big)\mathcal{A}^2+\big(2\varpi+3\big)r^2\big)^2}
\bigg\{\sqrt{2} \pi  \varpi  (\varpi +1) \\\nonumber
&\times\mathcal{A}^3 \big(\mathcal{A}^2+2 r^2\big)^2
(\mathcal{A}^2+2 \mathcal{C}^2) \tan ^{-1}\bigg(\frac{\sqrt{2}
r}{\sqrt{A}}\bigg)+2 r \big(\mathcal{A}^8 \big(6 (\varpi +2) \varpi
^2 \mathbb{C}_3 \mathcal{C}^2 \\\nonumber &+\pi \varpi (9 \varpi+36
(\varpi +2) \mathbb{C}_{3} \mathcal{C}^2+19)+24 \pi ^2 (\varpi +2)
(2 \mathbb{C}_{3} \mathcal{C}^2+1)\big)+2 \mathcal{A}^6 \\\nonumber
&\times\big((19 \varpi +37) \varpi ^2 \mathbb{C}_{3} r^2
\mathcal{C}^2+\pi \varpi \big(2 r^2 (7 (\varpi +2)+3 (19 \varpi +37)
\mathbb{C}_{3} \mathcal{C}^2)\\\nonumber &+(9 \varpi +19)
\mathcal{C}^2\big)+4 \pi ^2 \big(r^2 (9 \varpi +38 \varpi
\mathbb{C}_{3} \mathcal{C}^2+74 \mathbb{C}_{3} \mathcal{C}^2+17)+6
(\varpi +2) \\\nonumber &\times\mathcal{C}^2\big)\big)+2
\mathcal{A}^4 r^2 \big((46 \varpi +85) \varpi ^2 \mathbb{C}_{3} r^2
\mathcal{C}^2+\pi \varpi \big(r^2 (14 \varpi +6 (46 \varpi +85)
\\\nonumber &\times\mathbb{C}_{3} \mathcal{C}^2+27)+28 (\varpi +2)
\mathcal{C}^2\big)+4 \pi ^2 \big(r^2 (8 \varpi +2 (46 \varpi +85)
\mathbb{C}_{3} \mathcal{C}^2+15)\\\nonumber &+2 (9 \varpi +17)
\mathcal{C}^2\big)\big)+4 \mathcal{A}^2 r^4 \big(2 (13 \varpi +22)
\varpi ^2\mathbb{C}_{3} r^2 \mathcal{C}^2+\pi \varpi \big(2 r^2 (2
\varpi +6 \\\nonumber &\times (13 \varpi +22) \mathbb{C}_{3}
\mathcal{C}^2+3)+(14 \varpi +27) \mathcal{C}^2\big)+4 \pi ^2
\big(r^2 (2 \varpi +52 \varpi \mathbb{C}_{3}
\mathcal{C}^2\\\nonumber &+88 \mathbb{C}_{3} \mathcal{C}^2+3)+(8
\varpi +15) \mathcal{C}^2\big)\big)+8 (\varpi +2 \pi ) (2 \varpi +3)
r^6 \mathcal{C}^2 \big(3 \varpi \mathbb{C}_{3} r^2\\\label{60g} &+2
\pi \big(6 \mathbb{C}_{3}
r^2+1\big)\big)\big)\bigg\}\bigg],\\\nonumber
\tilde{P}_{r}&=\frac{1}{8\mathcal{C}^2\big(\varpi+2\pi\big)\big(\varpi+4\pi\big)}\bigg[\frac{\alpha\big(\mathcal{A}^2+3r^2\big)}
{8\pi\big((\varpi+2)\mathcal{A}^2+(2\varpi+3)r^2\big)}\bigg\{8(\varpi+2\pi)\\\nonumber
&\times(\varpi+4\pi)\mathbb{C}_{3}\mathcal{C}^2-\pi\bigg(\frac{\varpi\sqrt{2A}\tan^{-1}\big(\frac{\sqrt{2}r}{\sqrt{\mathcal{A}^2}}\big)}{r^3}
-\frac{4(3\varpi+8\pi)}{\mathcal{A}^2+2r^2}-\frac{2\varpi}{r^2}\bigg)\\\nonumber
&\times(\mathcal{A}^2+2\mathcal{C}^2)\bigg\}+\frac{1}{\big(\mathcal{A}^2+2r^2\big)^2}
\big\{2\varpi\big(-4\mathcal{A}^2r^2+3\mathcal{A}^2\mathcal{C}^2-6r^4+4r^2\mathcal{C}^2\big)\\\label{60h}
&-8\pi\big(\mathcal{A}^2+2r^2\big)\big(\mathcal{A}^2+3r^2-\mathcal{C}^2\big)\big\}\bigg],\\\nonumber
\tilde{P}_{\bot}&=\frac{1}{8\mathcal{C}^2\big(\varpi+2\pi\big)\big(\varpi+4\pi\big)}\bigg[
\frac{\alpha\big(\mathcal{A}^2+2r^2\big)}{16\pi
r^3\big(\mathcal{A}^2+r^2\big)\big((\varpi+2)\mathcal{A}^2+(2\varpi+3)r^2\big)^2}\\\nonumber
&\times\bigg\{\frac{2}{\big(\mathcal{A}^2+2r^2\big)^2}\big\{\mathcal{A}^8
r \big(8 \varpi ^2 (\varpi +2) \mathbb{C}_{3} r^2 \mathcal{C}^2+32
\pi ^2 (\varpi +2) r^2 (2 \mathbb{C}_{3} \mathcal{C}^2+1)\\\nonumber
&+\pi  \varpi \big(r^2 (5 \varpi +48 (\varpi +2) \mathbb{C}_{3}
\mathcal{C}^2+13)-2 (\varpi +2) \mathcal{C}^2\big)\big)-\pi  \varpi
(\varpi +2) \mathcal{A}^{10} r\\\nonumber &+\mathcal{A}^6 r^3
\big(48 (\varpi +2) \varpi ^2 \mathbb{C}_{3} r^2 \mathcal{C}^2+\pi
\varpi \big(r^2 (8 \varpi+288 (\varpi +2) \mathbb{C}_{3}
\mathcal{C}^2+27)\\\nonumber &+2 (5 \varpi +13)
\mathcal{C}^2\big)+64 \pi ^2 (\varpi +2) \big(r^2 (6 \mathbb{C}_{3}
\mathcal{C}^2+1)+\mathcal{C}^2\big)\big)+2 \mathcal{A}^4
r^5\big(4\varpi^2r^2\\\nonumber & \times (14 \varpi
+27)\mathbb{C}_{3} \mathcal{C}^2+\pi \varpi \big(r^2 (-6 \varpi +24
(14 \varpi +27) \mathbb{C}_{3}
\mathcal{C}^2-1)+\mathcal{C}^2\\\nonumber &\times(8 \varpi +27)
\big)+16 \pi ^2 \big(r^2 (28 \varpi \mathbb{C}_{3} \mathcal{C}^2+54
\mathbb{C}_{3} \mathcal{C}^2+1)+4 (\varpi +2)
\mathcal{C}^2\big)\big)\\\nonumber &+4 \mathcal{A}^2 r^7
\mathcal{C}^2 \big(8 (4 \varpi +7) \varpi ^2 \mathbb{C}_{3} r^2+\pi
\varpi \big(-6 \varpi +48 (4 \varpi +7) \mathbb{C}_{3}
r^2-1\big)\\\nonumber &+16 \pi ^2 \big(4 (4 \varpi +7)
\mathbb{C}_{3} r^2+1\big)\big)+32 (\varpi +2 \pi ) (\varpi +4 \pi )
(2 \varpi +3) \mathbb{C}_{3} r^{11} \mathcal{C}^2\big\}\\\nonumber
&+\pi\varpi\sqrt{2\mathcal{A}^2}(\mathcal{A}^2+2\mathcal{C}^2)
\big((\varpi+2)\mathcal{A}^4+(5\varpi+7)\mathcal{A}^2r^2+(2\varpi
+3)r^4\big)\\\nonumber
&\times\tan^{-1}\bigg(\frac{\sqrt{2}r}{\sqrt{\mathcal{A}^2}}\bigg)+\frac{\alpha
r^2 \big(2 \mathcal{A}^2+r^2\big)}{8 \pi \big(\mathcal{A}^2+r^2\big)
\big((\varpi+2)\mathcal{A}^2+(2\varpi+3)r^2\big)}\bigg\{8\mathbb{C}_{3}\mathcal{C}^2\\\nonumber
&\times(\varpi+2\pi)(\varpi+4\pi)-\pi\bigg(\frac{\varpi\sqrt{2A}\tan^{-1}\big(\frac{\sqrt{2}r}{\sqrt{\mathcal{A}^2}}\big)}{r^3}
-\frac{4(3\varpi+8\pi)}{\mathcal{A}^2+2r^2}-\frac{2\varpi}{r^2}\bigg)\\\nonumber
&\times(\mathcal{A}^2+2\mathcal{C}^2)\bigg\}\bigg\}+\frac{1}{\big(\mathcal{A}^2+2r^2\big)^2}
\big\{2\varpi\big(3\mathcal{A}^2\mathcal{C}^2-4\mathcal{A}^2r^2-6r^4+4r^2\mathcal{C}^2\big)\\\label{60i}
&-8\pi\big(\mathcal{A}^2+2r^2\big)
\big(\mathcal{A}^2+3r^2-\mathcal{C}^2\big)\big\}\bigg],
\end{align}
and the pressure anisotropy is given by
\begin{align}\nonumber
\tilde{\Pi}&=\frac{1}{128\pi\big(\varpi+2\pi\big)\big(\varpi+4\pi\big)r^3\mathcal{C}^2\big((\varpi+2)\mathcal{A}^2+(2\varpi+3)
r^2\big)^2}\bigg[\pi\sqrt{2\mathcal{A}^2}\\\nonumber &\times
\alpha\varpi\big\{3 (\varpi +2) \mathcal{A}^4+(12 \varpi +19)
\mathcal{A}^2 r^2+6 (2 \varpi +3) r^4\big\} \tan
^{-1}\bigg(\frac{\sqrt{2} r}{\sqrt{\mathcal{A}^2}}\bigg)\\\nonumber
&\times(\mathcal{A}^2+2 \mathcal{C}^2)-2 r\alpha \big\{3 \pi \varpi
(\varpi +2) \mathcal{A}^6+\pi\big(5 (2 \varpi +3) r^2+6 (\varpi +2)
\mathcal{C}^2\big)\\\nonumber &\times\varpi \mathcal{A}^4+2
\mathcal{A}^2 r^2 \big(-4 \varpi ^2 \mathbb{C}_{3} r^2
\mathcal{C}^2+16 \pi ^2 r^2 (2 \varpi -2 \mathbb{C}_{3}
\mathcal{C}^2+3)+\pi \varpi \big(5\mathcal{C}^2\\\label{g60j}
&\times(2 \varpi +3)+8 r^2 (2 \varpi -3 \mathbb{C}_{3}
\mathcal{C}^2+3)\big)\big)+32 \pi (\varpi +2 \pi ) (2 \varpi +3) r^4
\mathcal{C}^2\big\}\bigg].
\end{align}

\section{Graphical Interpretation of the Developed Solutions}

The mass of a spherical distribution is computed by numerically
solving the following differential equation
\begin{equation}\label{g63}
\frac{dm(r)}{dr}=4\pi r^2 \tilde{\mu},
\end{equation}
along with the initial condition $m(0)=0$. Here, $\tilde{\mu}$
represents the energy density in modified gravity corresponding to
each solution, whose value is given in Eqs.\eqref{g46} and
\eqref{60g}. The order in which intricate particles of
self-gravitating system are arranged, helps to measure the
compactness of that body $\big(\nu(r)\big)$. This factor determines
how tightly the particles in an object are packed. It is also gauged
by the mass-radius ratio of a stellar system. Buchdahl \cite{42a}
observed this factor as $\nu(r)<\frac{4}{9}$ in the case of
spherical spacetime. It is interestingly enough to know that the
compactness of a celestial object affects the wavelength of
neighboring electromagnetic radiations. The compact object having
sufficient gravitational attraction deviates the path of motion of
those waves from being straight. One can measure the redshift in
such radiations as
\begin{equation}
z(r)=\frac{1}{\sqrt{1-2\nu(r)}}-1,
\end{equation}
whose upper limits for perfect \cite{42a} and anisotropic
distributions \cite{42b} are $2$ and $5.211$, respectively.

In astrophysics, some constraints are very useful whose fulfillment
ensures the existence of normal (ordinary) matter in the interior of
compact structure, known as the energy conditions. These bounds also
confirm the physical viability of the fluid configuration. The
governing parameters of a geometry comprising of normal matter (such
as the energy density and pressure components) must obey these
constraints. In the current scenario, they turn out to be
\begin{eqnarray}\nonumber
&&\tilde{\mu} \geq 0, \quad \tilde{\mu}+\tilde{P}_{r} \geq
0,\\\nonumber &&\tilde{\mu}+\tilde{P}_{\bot} \geq 0, \quad
\tilde{\mu}-\tilde{P}_{r} \geq 0,\\\label{g50}
&&\tilde{\mu}-\tilde{P}_{\bot} \geq 0, \quad
\tilde{\mu}+\tilde{P}_{r}+2\tilde{P}_{\bot} \geq 0.
\end{eqnarray}
Another factor of great importance is the stability of a celestial
object. We firstly examine this phenomenon through the radial
$\big(v_{sr}^{2}=\frac{d\tilde{P}_{r}}{d\tilde{\mu}}\big)$ as well
as tangential
$\big(v_{s\bot}^{2}=\frac{d\tilde{P}_{\bot}}{d\tilde{\mu}}\big)$
sound speeds. The causality would be maintained if the sound speed
is less than the speed of light in the considered medium, i.e., $0 <
v_{sr}^{2},~ v_{s\bot}^{2} < 1$ \cite{42bb}. The stability can also
be confirmed by Herrera cracking concept, which states that a stable
system must fulfill the inequality $0 < |v_{s\bot}^{2}-v_{sr}^{2}| <
1$ \cite{42ba}. We also explore stability of the resulting models by
the adiabatic index $(\Gamma)$. According to this, the system shows
stable behavior only if $\Gamma > \frac{4}{3}$ \cite{42c}. We
express $\tilde{\Gamma}$ in this case as
\begin{equation}\label{g62}
\tilde{\Gamma}=\frac{\tilde{\mu}+\tilde{P}_{r}}{\tilde{P}_{r}}
\bigg(\frac{d\tilde{P}_{r}}{d\tilde{\mu}}\bigg).
\end{equation}
\begin{figure}\center
\epsfig{file=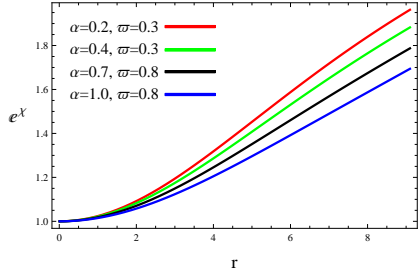,width=0.4\linewidth} \caption{Plot
of deformed radial metric \eqref{g40a} for the solution
corresponding to $\tilde{\Pi}=0$.}
\end{figure}

We consider $f(\mathbb{R},\mathbb{T})$ model \eqref{g61} to
interpret both the obtained solutions, deformation functions and the
complexity factor graphically. For this purpose, we choose multiple
values of the coupling and decoupling parameters along with
$\mathbb{C}_{1}=-0.003$, and explore different physical features of
the considered compact star. Figure \textbf{1} exhibits plot of the
deformed radial metric potential \eqref{g40a} and we observe its
non-singular and increasing behavior for $0<r<\mathcal{R}$. The
acceptability criteria of any gravitational model requires that the
governing parameters, representing fluid distribution (such as
energy density and pressure), must be maximum and finite in the core
of astrophysical body, and monotonically decreasing towards its
boundary. Figure \textbf{2} contains plots of the solution
\eqref{g46}-\eqref{g49} and we observe its acceptable behavior. The
energy density (left upper plot) is maximum in the middle and
decreases with the increment in both the coupling and decoupling
parameters. However, the pressure components show counter behavior
as they increase by increasing $\alpha$ as well as $\varpi$. The
radial pressure disappears at the boundary for all the considered
values of these parameters. We observe from Figure \textbf{2} (last
plot) that pressure anisotropy is zero at the center for
$\alpha=0.2,~0.4,~0.7$ and vanishes throughout for $\alpha=1$ which
confirms the system to be isotropic at this point.
\begin{figure}\center
\epsfig{file=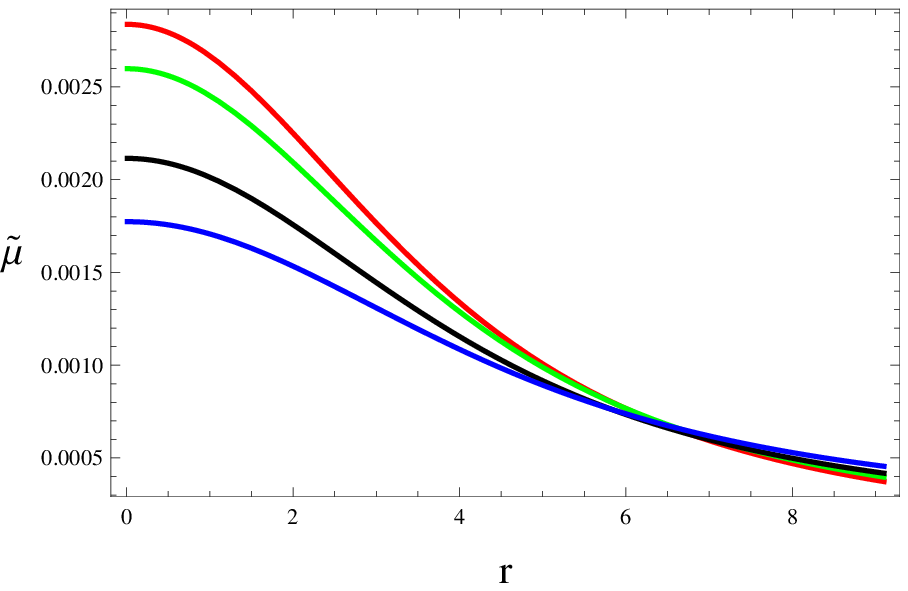,width=0.4\linewidth}\epsfig{file=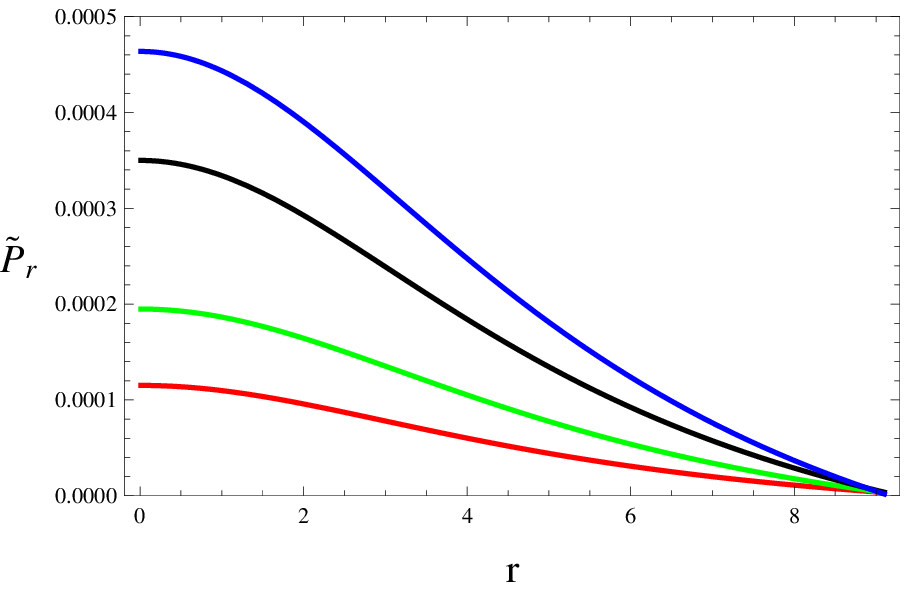,width=0.4\linewidth}
\epsfig{file=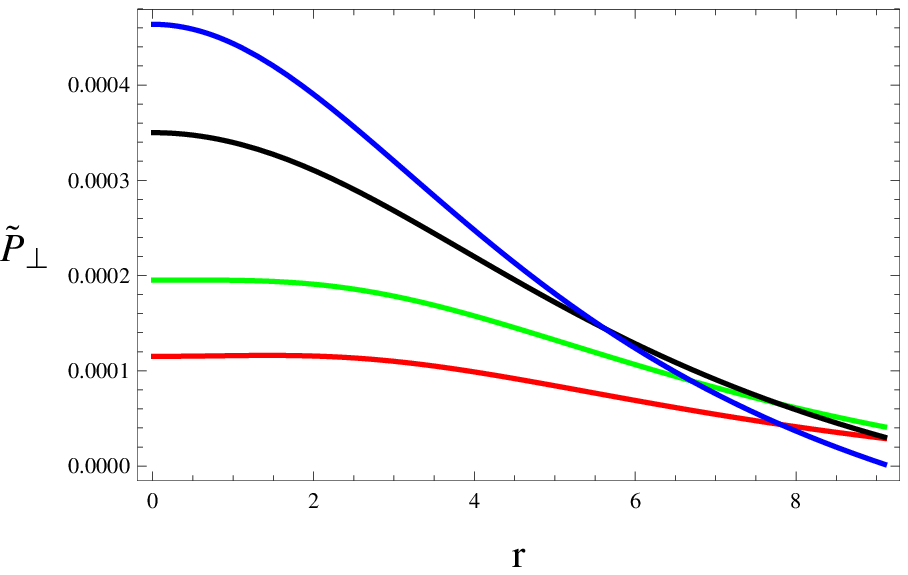,width=0.4\linewidth}\epsfig{file=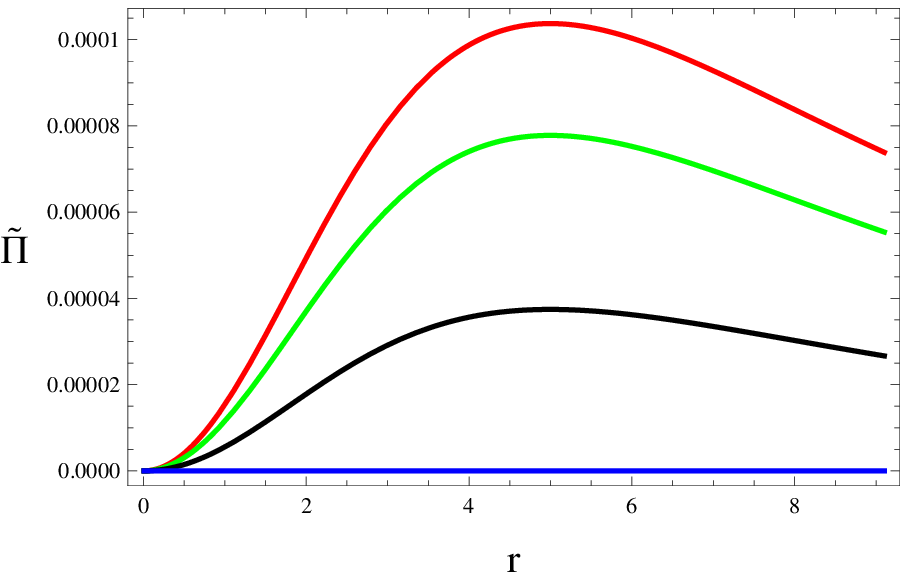,width=0.4\linewidth}
\caption{Plots of matter variables and anisotropy for the solution
corresponding to $\tilde{\Pi}=0$.}
\end{figure}
\begin{figure}\center
\epsfig{file=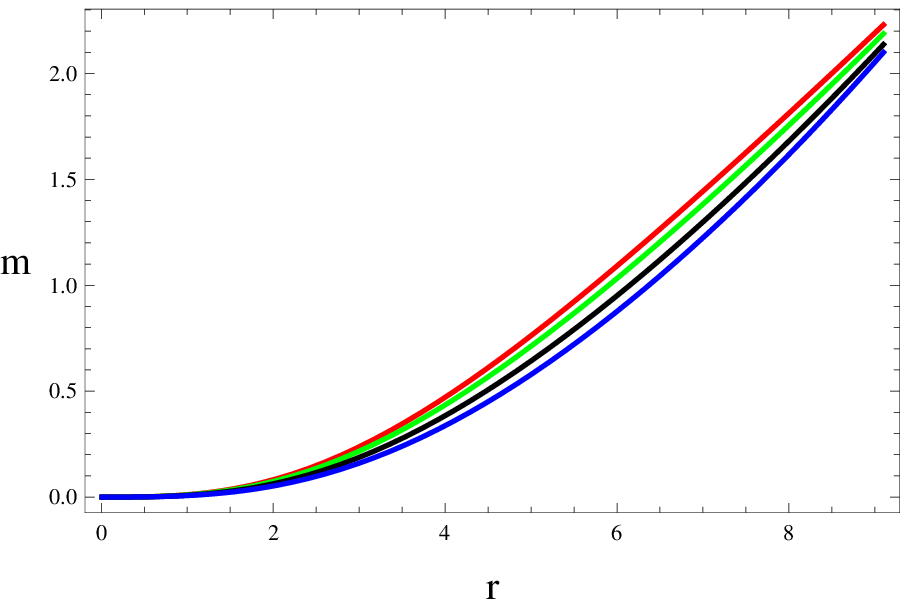,width=0.4\linewidth}\epsfig{file=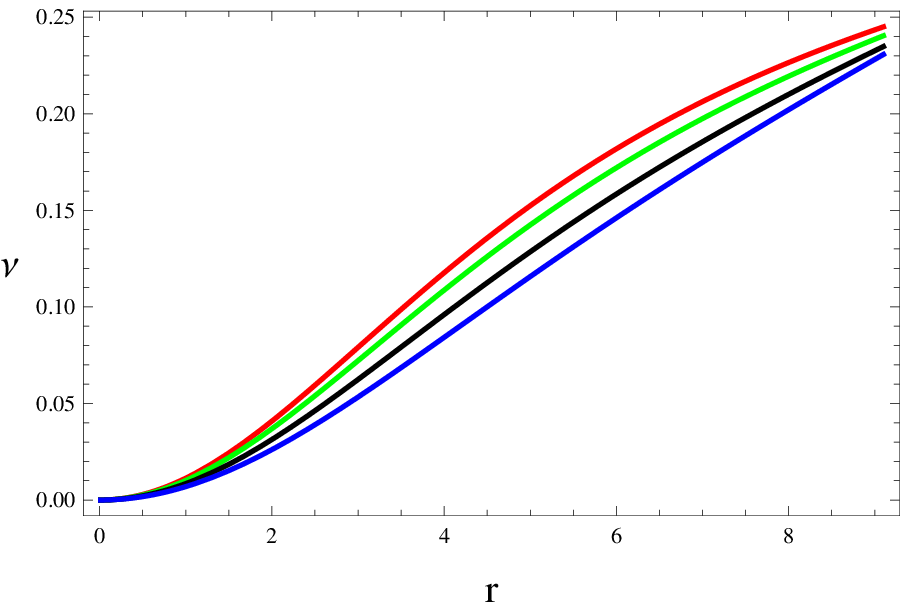,width=0.4\linewidth}
\epsfig{file=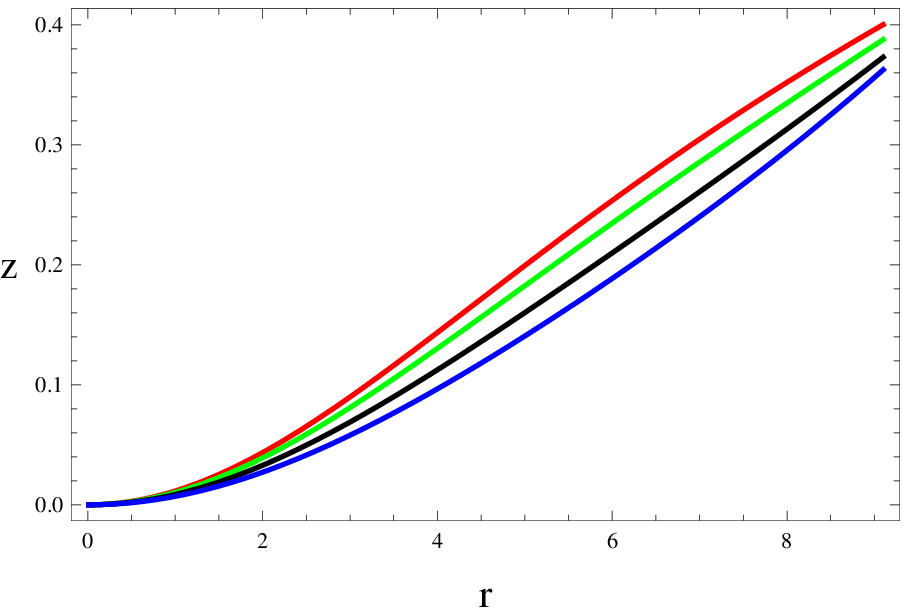,width=0.4\linewidth} \caption{Plots of
mass, compactness and redshift for the solution corresponding to
$\tilde{\Pi}=0$.}
\end{figure}
\begin{figure}\center
\epsfig{file=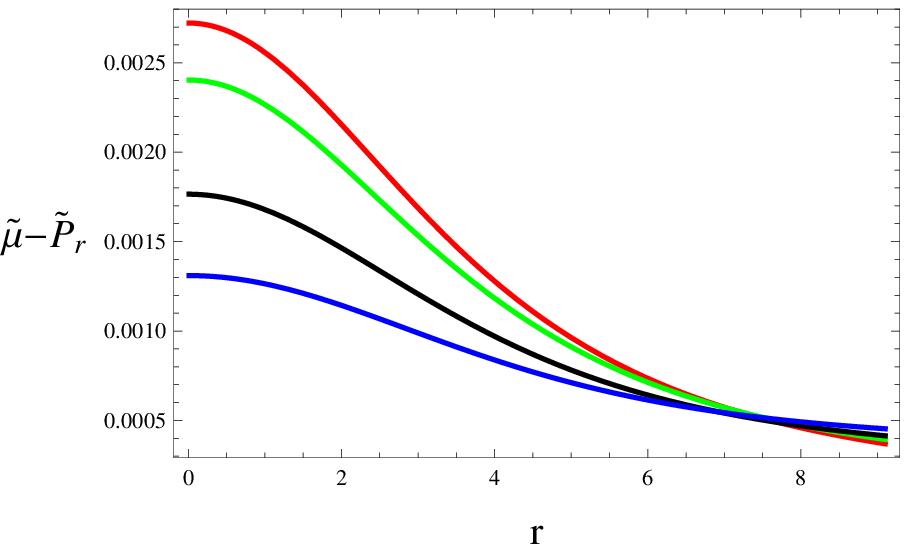,width=0.4\linewidth}\epsfig{file=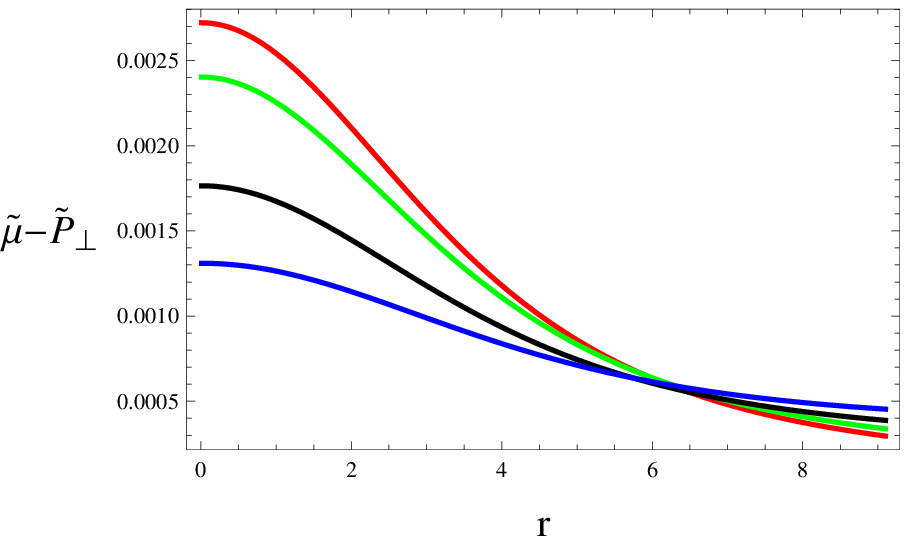,width=0.4\linewidth}
\caption{Plots of dominant energy conditions for the solution
corresponding to $\tilde{\Pi}=0$.}
\end{figure}
\begin{figure}\center
\epsfig{file=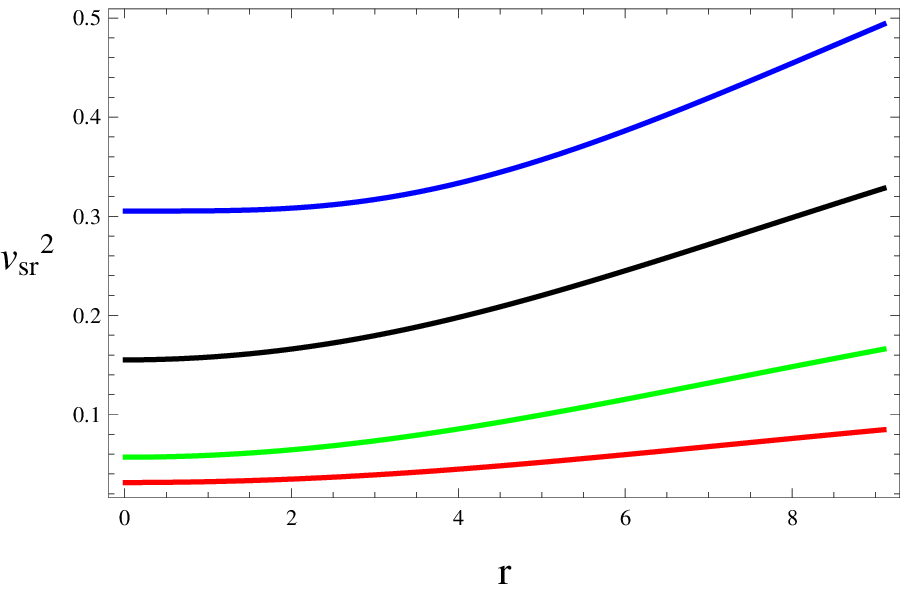,width=0.4\linewidth}\epsfig{file=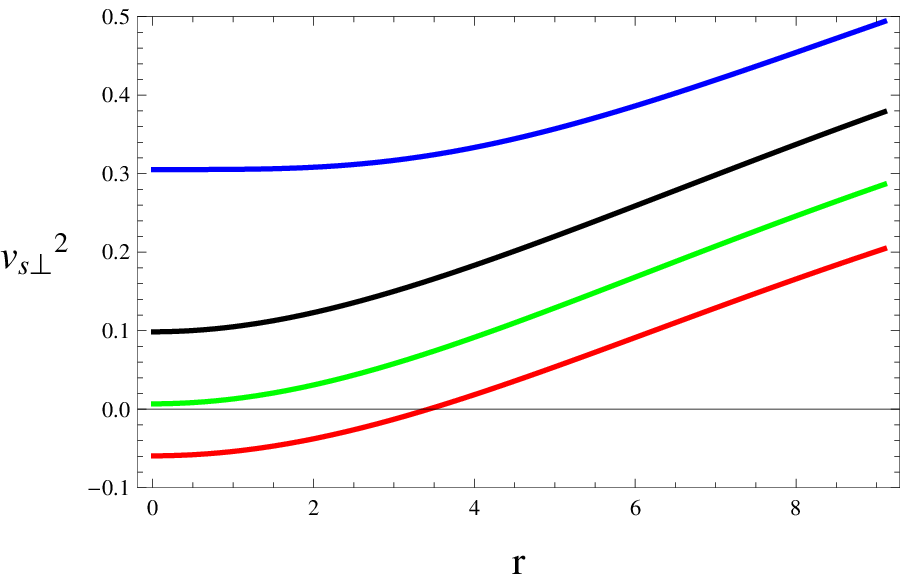,width=0.4\linewidth}
\epsfig{file=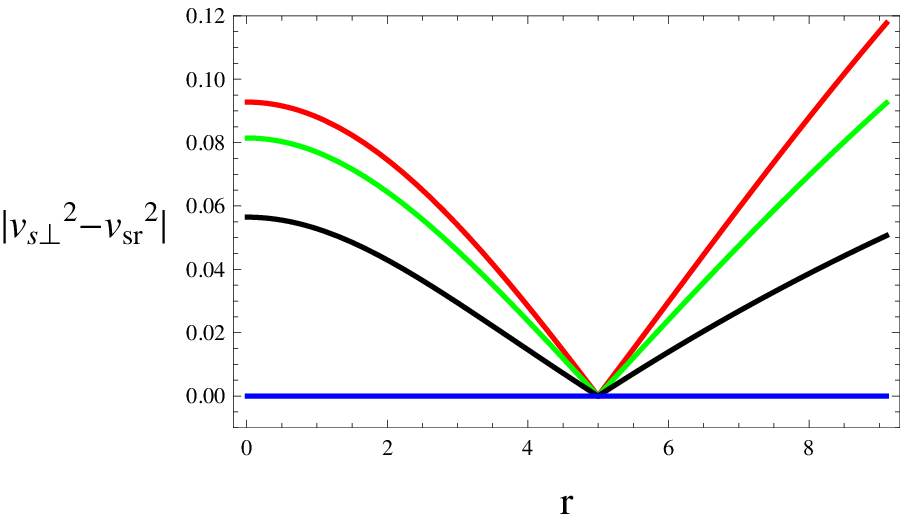,width=0.4\linewidth}\epsfig{file=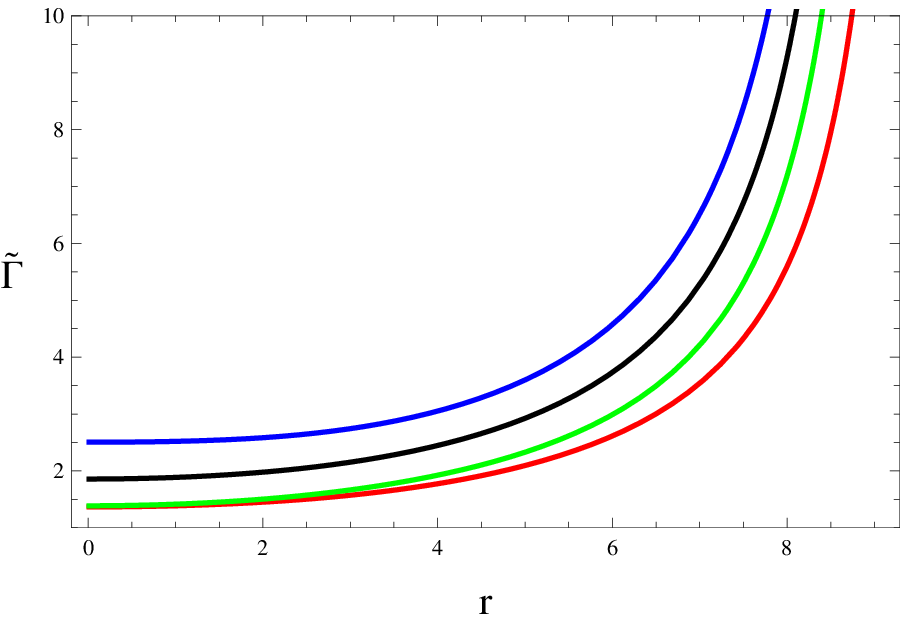,width=0.4\linewidth}
\caption{Plots of radial/tangential
velocities,~$|v_{s\bot}^2-v_{sr}^2|$ and adiabatic index for the
solution corresponding to $\tilde{\Pi}=0$.}
\end{figure}

The mass of spherical geometry is presented in Figure \textbf{3},
from which we observe that anisotropic system is more massive and
dense as compared to isotropic analog. The other two plots also
confirm the fulfillment of required limits of both the compactness
and redshift. The state variables show positive trend, thus we only
need to plot dominant energy conditions as
$\tilde{\mu}-\tilde{P}_{r} \geq 0$ and $\tilde{\mu}-\tilde{P}_{\bot}
\geq 0$. Figure \textbf{4} reveals the viability of our resulting
solution as these bounds are satisfied. The stability is checked in
Figure \textbf{5} through different approaches. According to the
sound speed, the system is unstable near the core for $\alpha=0.2$
as $v_{s\bot}^{2} < 0$, and stable everywhere for all other values
of this parameter. However, Herrera's cracking approach and the
adiabatic index ensure the stability of spherical structure for all
choices of $\alpha$ and $\varpi$ (lower two plots). The complexity
factors \eqref{g56a} and \eqref{g60d} are plotted in Figure
\textbf{6}, and we notice that they decrease with the increment in
coupling and decoupling parameters. This follows that
$f(\mathbb{R},\mathbb{T})$ theory reduces the impact of complexity
as compared to $\mathbb{GR}$.
\begin{figure}\center
\epsfig{file=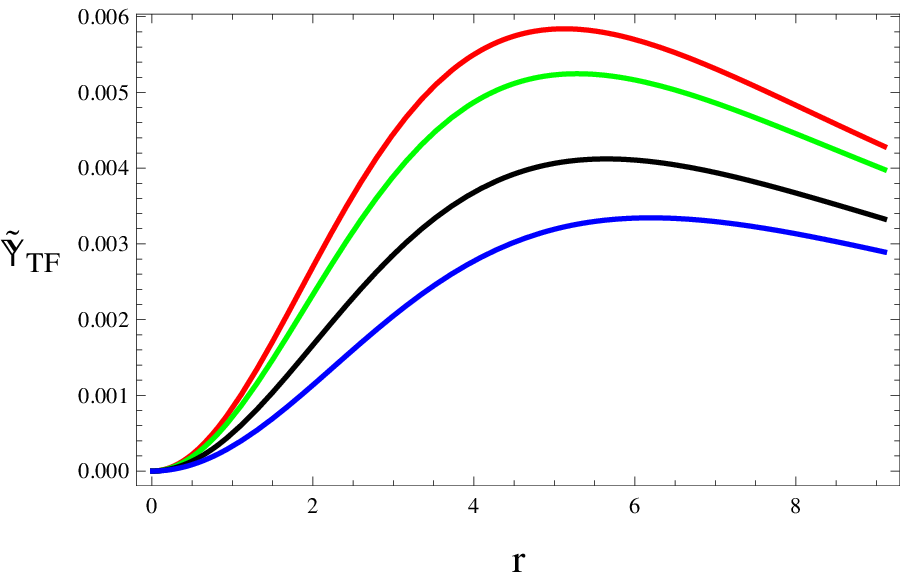,width=0.4\linewidth}\epsfig{file=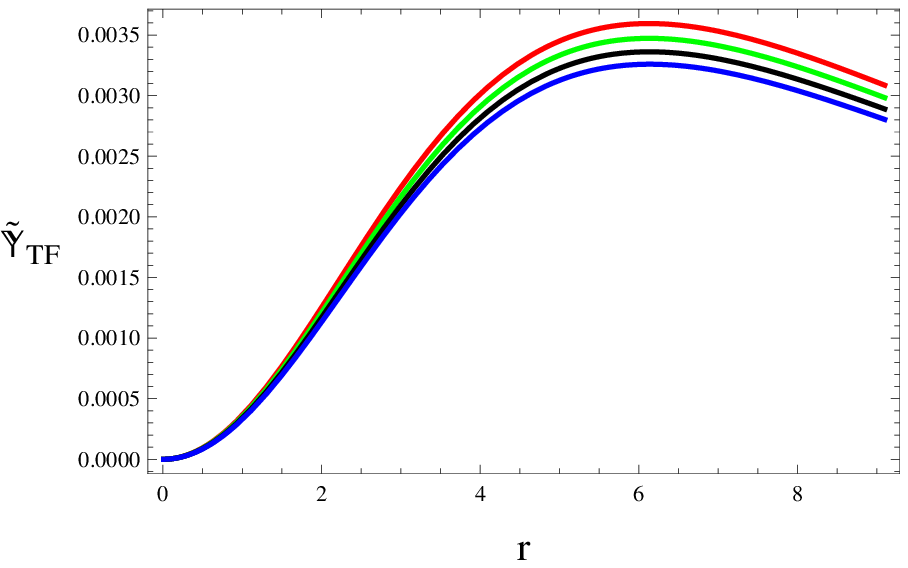,width=0.4\linewidth}
\caption{Plots of complexity factors \eqref{g56a} and \eqref{g60d}.}
\end{figure}

We now explore physical characteristics of the solution
corresponding to $\tilde{\mathbb{Y}}_{TF}=0$ by choosing
$\mathbb{C}_{3}=-0.001$. The nature of deformed radial component is
also found to be non-singular, as shown in Figure \textbf{7}. Figure
\textbf{8} demonstrates the plots of matter variables
\eqref{60g}-\eqref{60i} and anisotropic pressure \eqref{g60j}. They
show the same behavior as we have found in the previous solution.
The anisotropic factor vanishes at the center of star and then show
negative trend towards the boundary, i.e., $\tilde{P}_\bot <
\tilde{P}_r$. Figure \textbf{9} exhibits mass of the spherical
geometry that decreases for higher values of parameters $\alpha$ and
$\varpi$. The compactness and redshift also meet the required
criteria (right and lower plots). The dominant energy conditions are
plotted in Figure \textbf{10} whose fulfillment confirms viability
of the corresponding solution as well as modified model \eqref{g61}.
All the plots in Figure \textbf{11} reveal that our developed
solution \eqref{60g}-\eqref{g60j} is stable everywhere.
\begin{figure}\center
\epsfig{file=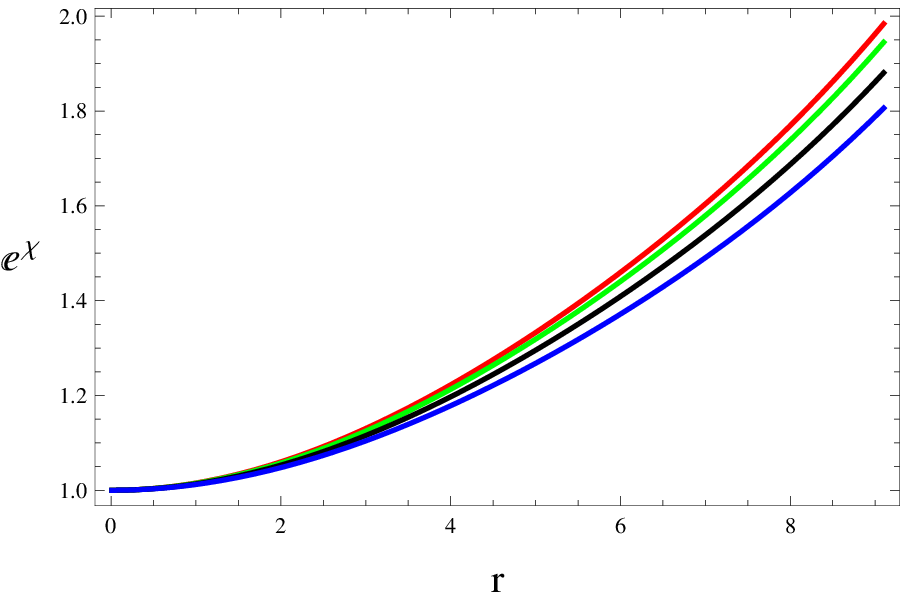,width=0.4\linewidth} \caption{Plot
of deformed radial metric \eqref{g60fa} for the solution
corresponding to $\tilde{\mathbb{Y}}_{TF}=0$.}
\end{figure}
\begin{figure}\center
\epsfig{file=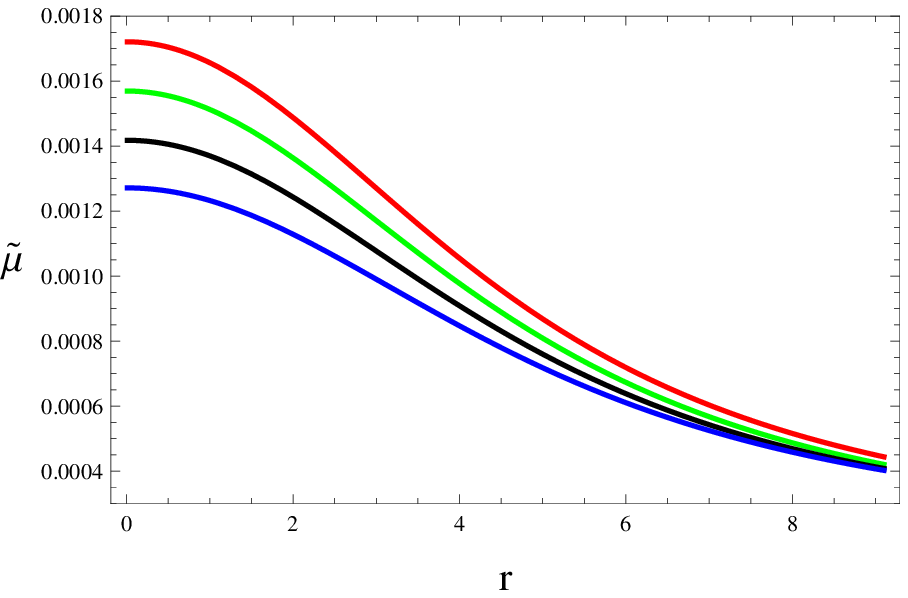,width=0.4\linewidth}\epsfig{file=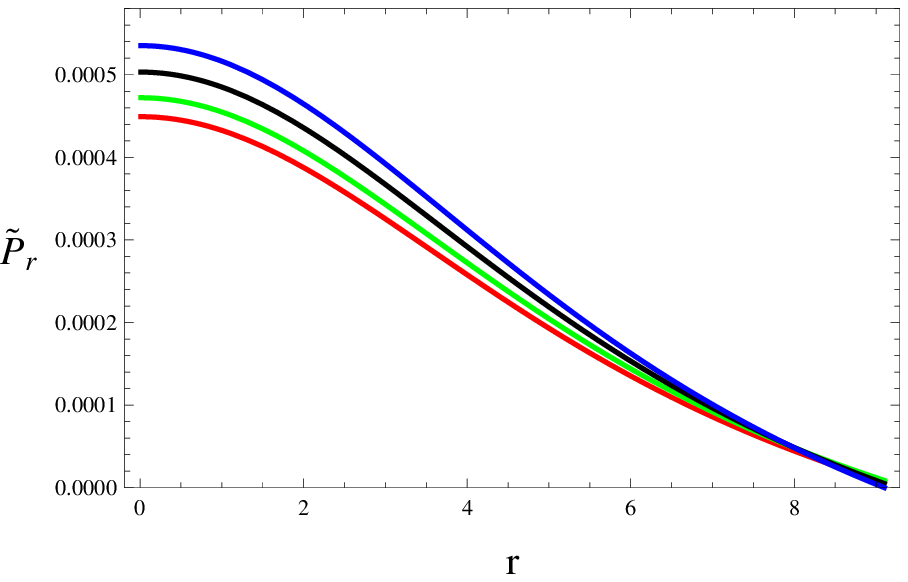,width=0.4\linewidth}
\epsfig{file=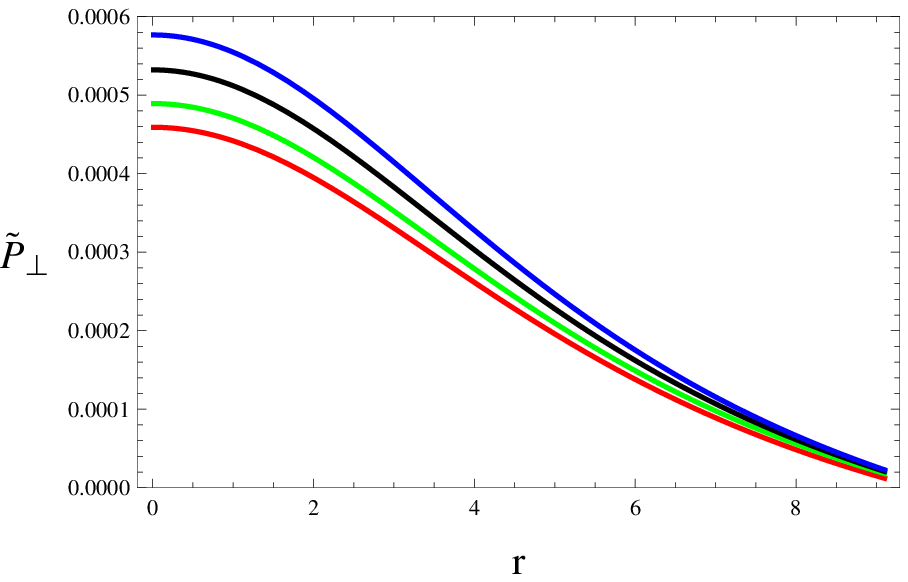,width=0.4\linewidth}\epsfig{file=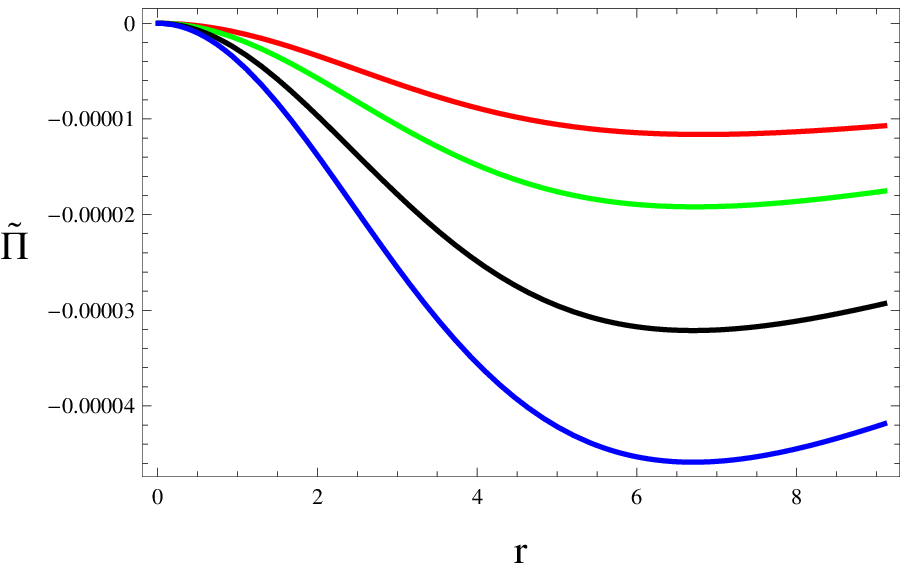,width=0.4\linewidth}
\caption{Plots of matter variables and anisotropy for the solution
corresponding to $\tilde{\mathbb{Y}}_{TF}=0$.}
\end{figure}
\begin{figure}\center
\epsfig{file=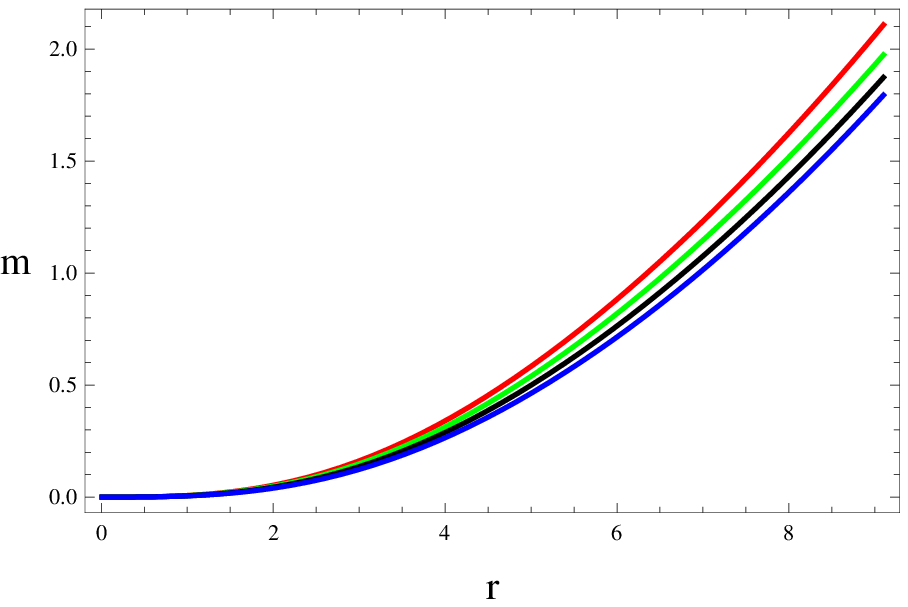,width=0.4\linewidth}\epsfig{file=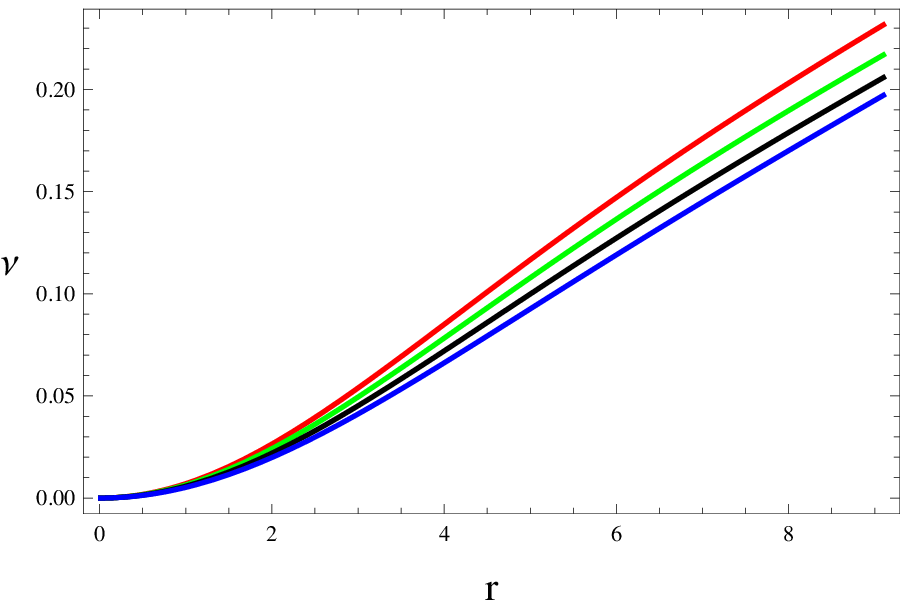,width=0.4\linewidth}
\epsfig{file=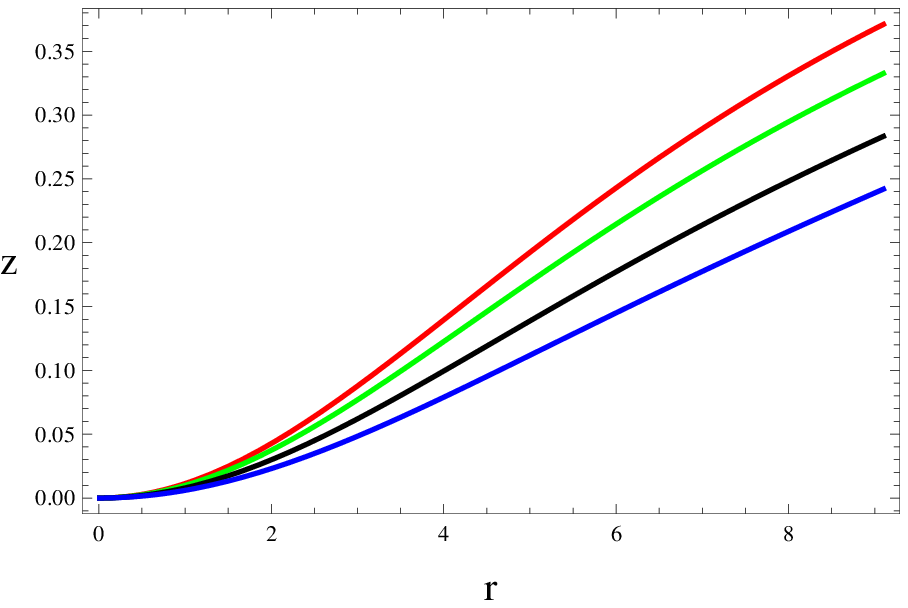,width=0.4\linewidth} \caption{Plots of
mass, compactness and redshift for the solution corresponding to
$\tilde{\mathbb{Y}}_{TF}=0$.}
\end{figure}
\begin{figure}\center
\epsfig{file=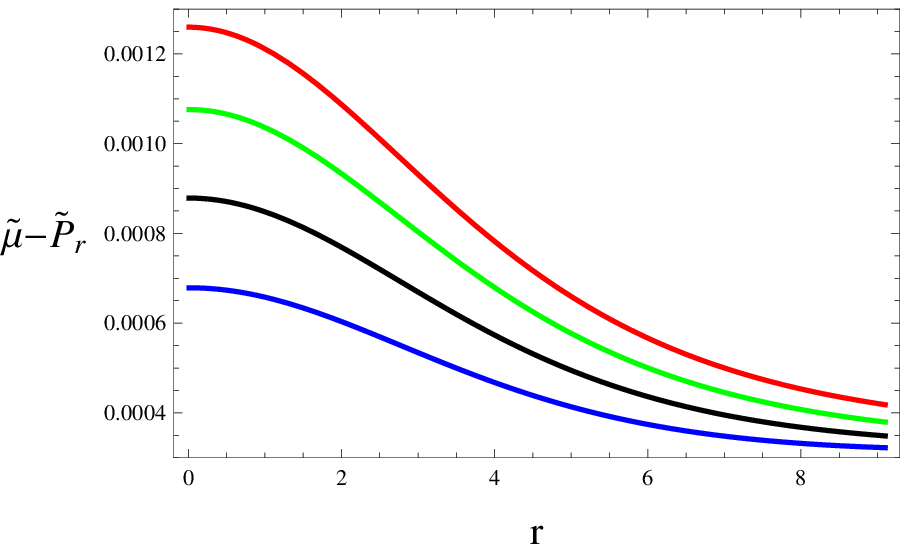,width=0.4\linewidth}\epsfig{file=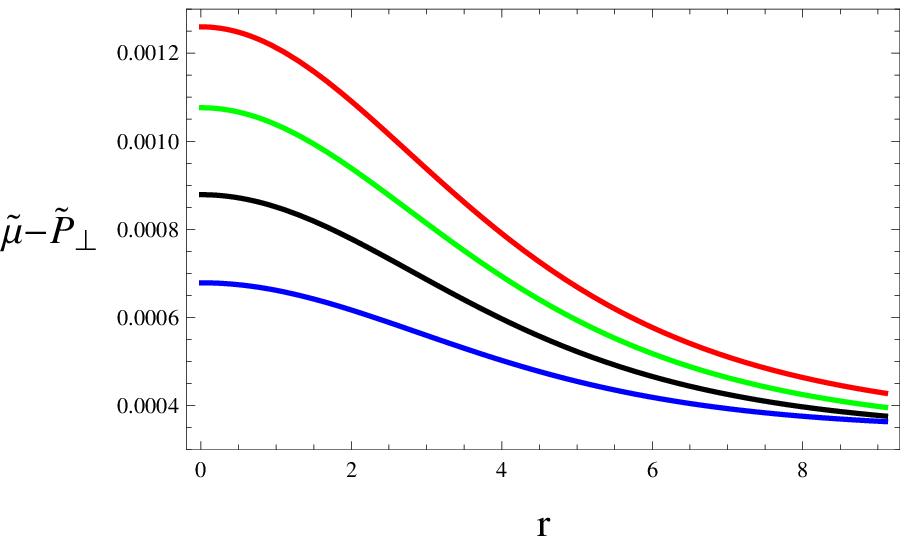,width=0.4\linewidth}
\caption{Plots of dominant energy conditions for the solution
corresponding to $\tilde{\mathbb{Y}}_{TF}=0$.}
\end{figure}
\begin{figure}\center
\epsfig{file=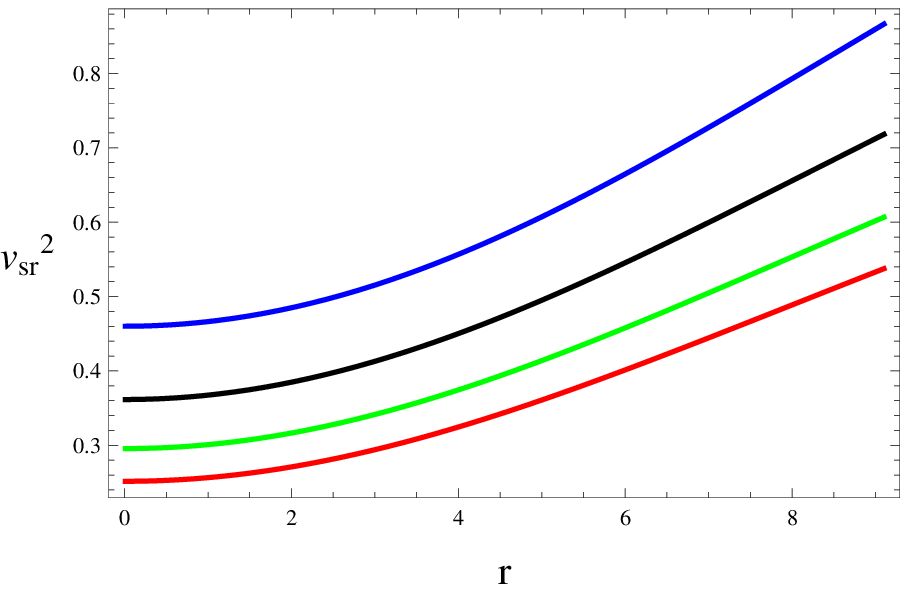,width=0.4\linewidth}\epsfig{file=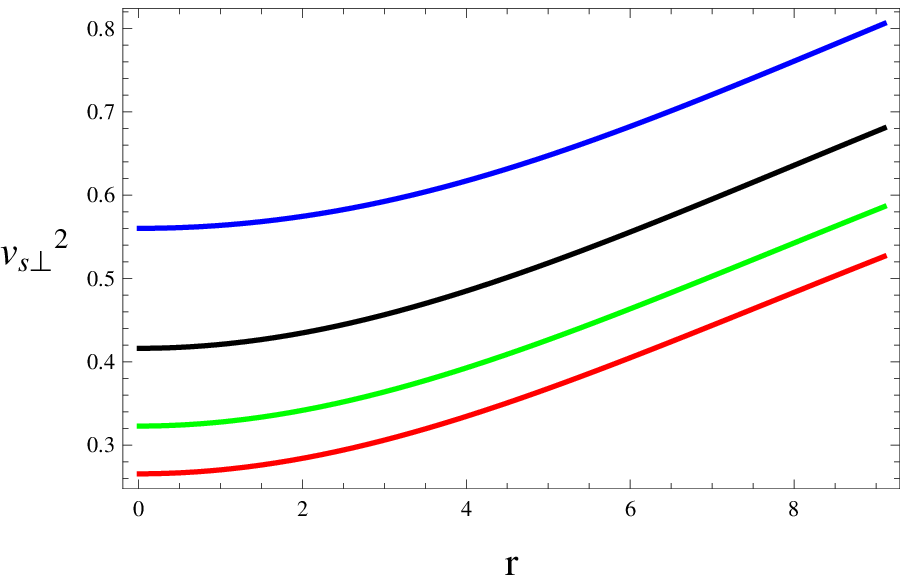,width=0.4\linewidth}
\epsfig{file=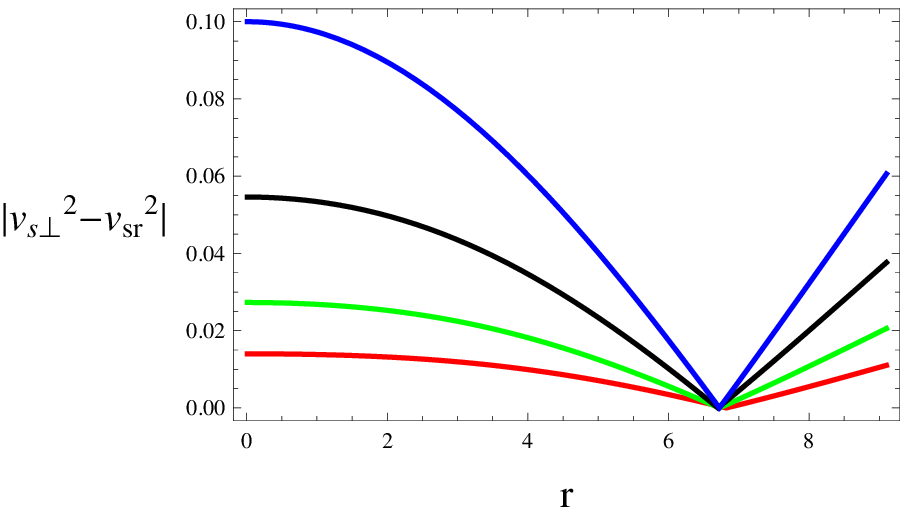,width=0.4\linewidth}\epsfig{file=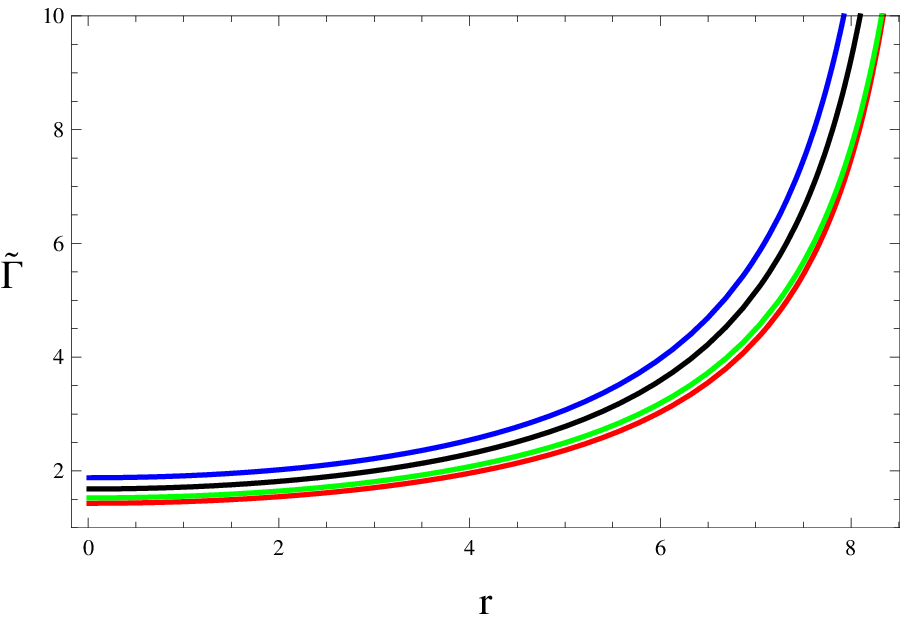,width=0.4\linewidth}
\caption{Plots of radial/tangential
velocities,~$|v_{s\bot}^2-v_{sr}^2|$ and adiabatic index for the
solution corresponding to $\tilde{\mathbb{Y}}_{TF}=0$.}
\end{figure}

\section{Conclusions}

In this paper, we have extended the existing solutions corresponding
to self-gravitating anisotropic sphere by adding an extra source
with the help of gravitational decoupling in
$f(\mathbb{R},\mathbb{T})=\mathbb{R}+2\varpi\mathbb{T}$ gravity. We
have formulated the modified field equations comprising the effects
of both sources and then separated them into two sets through the
MGD technique. Both the obtained sectors correspond to the original
anisotropic and the additional source, respectively. To deal with
the first set, we have used
$$\sigma(r)=\ln\bigg\{\mathcal{B}^2\bigg(1+\frac{r^2}{\mathcal{A}^2}\bigg)\bigg\},\quad
\xi(r)=e^{-\chi(r)}=\frac{\mathcal{A}^2+r^2}{\mathcal{A}^2+3r^2},$$
and metric potentials of Tolman IV ansatz, leading to two different
solutions. The unknowns involving in these solutions have been
calculated through boundary conditions for the mass and radius of
$4U 1820-30$. The second sector \eqref{g21}-\eqref{g23} contains
four unknowns, thus we have implemented extra constraints on the
newly added source $\mathfrak{A}_{\zeta\eta}$. We have considered
vanishing of the effective anisotropy to obtain the first solution
which leads to an isotropic system for $\alpha=1$. The other
solution is obtained by taking into the account that complexity of
the original and additional matter sources cancel out the effect of
each other.

The physical features of the obtained results have been analyzed by
taking $\alpha=0.2,~0.4,~0.7,~1$ and $\varpi=0.3,~0.8$. The
acceptable behavior of the corresponding state variables
\big(\eqref{g46}-\eqref{g48} and \eqref{60g}-\eqref{60i}\big),
anisotropy \big(\eqref{g49} and \eqref{g60j}\big) and the energy
conditions \eqref{g50} have been observed for specific values of the
integration constants. We have also found fulfillment of the
required limit for redshift and compactness (Figures \textbf{3} and
\textbf{9}). It is noticed that the solution corresponding to
$\tilde{\Pi}=0$ produces more dense stellar structure for all values
of $\alpha$ and $\varpi$, as compared to the other solution. The
deformation functions \eqref{g40} and \eqref{60f} are zero at the
center and exhibit positive behavior throughout. We have checked
stability of both the extended solutions through multiple approaches
such as the sound speed, cracking approach and the adiabatic index,
and their corresponding criteria is fulfilled, hence these solutions
are stable. It must be mentioned that our solutions are consistent
with $\mathbb{GR}$ \cite{37k}. Moreover, the solution corresponding
to $\tilde{\Pi}=0$ shows compatible behavior with the Brans-Dicke
gravity \cite{37m}, as it shows unstable behavior for $\alpha=0.2$
in that case as well (Figure \textbf{5}). Finally, all of our
results reduce to $\mathbb{GR}$ for $\varpi=0$.\\\\
\textbf{Data Availability Statement:} This manuscript has no
associated data.

\end{document}